\pdfoutput=1

\documentclass[sigconf]{acmart}

\AtBeginDocument{%
  \providecommand\BibTeX{{%
    \normalfont B\kern-0.5em{\scshape i\kern-0.25em b}\kern-0.8em\TeX}}}

\setcopyright{acmcopyright}
\copyrightyear{2018}
\acmYear{2018}
\acmDOI{XXXXXXX.XXXXXXX}

\usepackage{color}

\usepackage[utf8]{inputenc} 
\usepackage[T1]{fontenc}    
\usepackage{hyperref}       
\usepackage{url}            
\usepackage{booktabs}       
\usepackage{amsfonts}       
\usepackage{nicefrac}       
\usepackage{microtype}      
\usepackage{xcolor}         
\usepackage{float}
\usepackage{booktabs}
\usepackage{array}
\usepackage{balance} 
\usepackage{multirow}
\usepackage[normalem]{ulem}
\usepackage{color}
\definecolor{lightgray}{RGB}{215,215,215}
\usepackage{colortbl}  
\usepackage{xcolor}
\useunder{\uline}{\ul}{}
\usepackage{subfigure}

\usepackage{algorithm}  
\usepackage{algorithmicx}  
\usepackage{algpseudocode}  
\usepackage{amsmath}  
\usepackage{enumitem}

\floatname{algorithm}{Algorithm}

\usepackage{amsmath}  
\usepackage{enumitem}
\usepackage{tabularx}
\usepackage[utf8]{inputenc}
\usepackage[english]{babel}
\usepackage{amsthm}
\usepackage{bm}
\usepackage{graphicx}
\usepackage{caption}
\usepackage{wrapfig}
\usepackage{verbatim}

\newtheorem{assumption}{Assumption}[section]
\newcommand{\ie}{\emph{i.e., }}
\newcommand{\eg}{\emph{e.g., }}

\newcommand{\wrt}{\emph{w.r.t. }}

\newcommand{\aka}{\emph{a.k.a. }}

\newtheorem{proposition}{Proposition}

\copyrightyear{2024}
\acmYear{2024}
\setcopyright{acmlicensed}\acmConference[WWW '24]{Proceedings of the ACM Web Conference 2024}{May 13--17, 2024}{Singapore, Singapore}
\acmBooktitle{Proceedings of the ACM Web Conference 2024 (WWW '24), May 13--17, 2024, Singapore, Singapore}
\acmDOI{10.1145/3589334.3645403}
\acmISBN{979-8-4007-0171-9/24/05}

\begin{document}

\title{Uplift Modeling for Target User Attacks on Recommender Systems}

\author{Wenjie Wang}
\authornote{The first two authors contributed equally to this research.}
\email{wenjiewang96@gmail.com}
\orcid{0000-0002-5199-1428}
\affiliation{%
  \institution{National University of Singapore}
  \country{Singapore}
}

\author{Changsheng Wang}
\authornotemark[1]
\email{wcsa23187@gmail.com}
\orcid{0009-0007-0957-638X}
\affiliation{%
  \institution{University of Science and Technology of China}
  \city{Hefei}
  \state{Anhui}
  \country{China}
}

\author{Fuli Feng}
\email{fulifeng93@gmail.com}
\authornote{Corresponding author. This work is supported by the National Key Research and Development Program of China (2022YFB3104701), the National Natural Science Foundation of China (62272437), and the CCCD Key Lab of Ministry of Culture and Tourism.}
\orcid{0000-0002-5828-9842}
\affiliation{%
  \institution{University of Science and Technology of China}
  \city{Hefei}
  \state{Anhui}
  \country{China}
}

\author{Wentao Shi}
\email{shiwentao123@mail.ustc.edu.cn}
\orcid{0000-0002-2616-6880}
\affiliation{%
  \institution{University of Science and Technology of China}
  \city{Hefei}
  \state{Anhui}
  \country{China}
}

\author{Daizong Ding}
\email{17110240010@fudan.edu.cn}
\orcid{0000-0002-4722-5229}
\affiliation{%
  \institution{Fudan University}
  \city{Shanghai}
  \country{China}
}

\author{Tat-Seng Chua}
\email{dcscts@nus.edu.sg}
\orcid{0000-0001-6097-7807}
\affiliation{%
  \institution{National University of Singapore}
  \country{Singapore}
}



\begin{abstract}
Recommender systems are vulnerable to injective attacks, which inject limited fake users into the platforms to manipulate the exposure of target items to all users. In this work, we identify that conventional injective attackers overlook the fact that each item has its unique potential audience, and meanwhile, the attack difficulty across different users varies. Blindly attacking all users will result in a waste of fake user budgets and inferior attack performance. 
To address these issues, we focus on an under-explored attack task called target user attacks, aiming at promoting target items to a particular user group. In addition, we formulate the varying attack difficulty as heterogeneous treatment effects through a causal lens and propose an Uplift-guided Budget Allocation (UBA) framework. UBA estimates the treatment effect on each target user and optimizes the allocation of fake user budgets to maximize the attack performance. 
Theoretical and empirical analysis demonstrates the rationality of treatment effect estimation methods of UBA. 
By instantiating UBA on multiple attackers, we conduct extensive experiments on three datasets under various settings with different target items, target users, fake user budgets, victim models, and defense models, validating the effectiveness and robustness of UBA. 


\vspace{-0.1cm}
\end{abstract}





\maketitle

\section{Introduction}
\label{sec:introduction}

\begin{figure}[t]
\setlength{\abovecaptionskip}{0.1cm}
\setlength{\belowcaptionskip}{-0.3cm}
\centering
\includegraphics[scale=0.6]{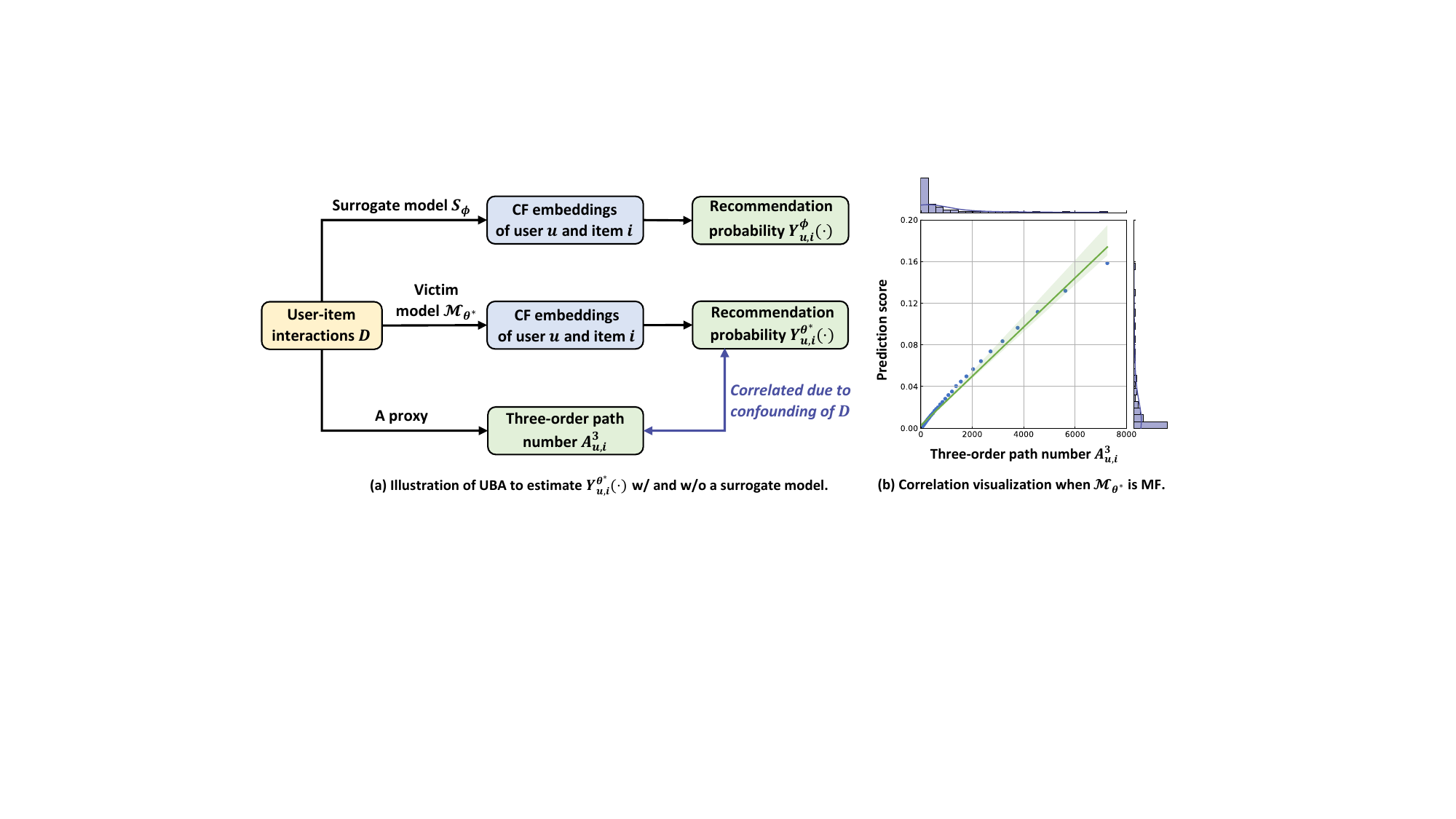}
\caption{Illustration of injective attacks on all users and target users (a) and varying attack difficulty on two users (b), where one fake user may cause different uplifts of the recommendation probabilities.}
\label{fig:target_users}
\end{figure}

Recommender systems have evolved into fundamental services for information filtering on numerous Web platforms such as Amazon and Twitter. 
Recent research has validated the vulnerability of recommender systems to \textit{injective attacks}~\cite{li2022revisiting,zhu2013differential,zhang2020adversarial}, which aim to promote the exposure of a target item via injecting limited fake users (refer to Figure~\ref{fig:target_users}(a)). 
Specifically, since recommender models typically utilize Collaborative Filtering (CF) in users' historical interactions to make recommendations, the attackers can fabricate fake user interactions and inject them into open-world Web platforms, so as to induce recommender models to elevate the exposure probability of a target item. As a result, such injective attacks can deliberately amplify traffic to target items, potentially bringing economic, political, or other profits to certain entities. 

Generally, past literature on injective attacks falls into three main groups: 
1) \textit{Heuristic attackers}~\cite{burke2005limited,kaur2016shilling} that adopt heuristic rules to construct fake users, for instance, Bandwagon Attack~\cite{1565730} increases the co-occurrence of popular items and the target item in fake user interactions; 
2) \textit{gradient-based attackers}~\cite{li2016data,fang2020influence} that directly adjust the interactions of fake users via gradients to maximize the well-designed attack objectives; 
and 3) \textit{neural attackers}~\cite{tang2020revisiting,Lin_2022,lin2020attacking} that optimize the neural networks to generate influential fake users for promoting the target item to more users.



However, previous work neglects that not all users will be interested in the target item, as each item appeals to its unique audience. For instance, dresses typically appeal more to female buyers. Increasing the recommendations of a target item to all users not only wastes attack resources, but also results in inferior attack performance. 
Worse still, most studies fail to account for the varying attack difficulty across different users~\cite{li2022revisiting} (see Figure~\ref{fig:target_users}(b)). Some ``harder'' users need more fake users to receive exposure to a target item. 
Due to ignoring the varying attack difficulty, easy users might receive redundant fake user budgets while hard users might be inadequately attacked, leading to inefficient use of budgets and poor attack performance.

To address the issues, we focus on an interesting recommender attack task --- \textit{target user attacks}, which attempt to expose a target item to a specific user group instead of all users. Moreover, we formulate the varying attack difficulty via causal language. From a causal view, assigning fake user budgets to a target user can be formulated as a treatment, and the probability of recommending a target item to this target user is the outcome. Given a target item, the varying attack difficulty essentially reflects the heterogeneous treatment effects (\aka uplifts\footnote{Treatment effects and uplifts are exchangeable in this work~\cite{Ai-LBCF-2022,zhang_unified_2021}
.}) on different target users. This is attributed to the different similarities between the target item and each target user's historically liked items. In this light, the key to maximizing attack performance with limited fake user budgets lies in: 1) estimating the heterogeneous treatment effect on each target user with different budgets, and 2) allocating the limited budgets wisely to maximize the treatment effects across all target users, \ie the overall recommendation probability.

To this end, we present an Uplift-guided Budget Allocation (UBA) framework for target user attacks. In particular, UBA utilizes two methods to estimate the treatment effect. If a surrogate recommender model (\eg Matrix Factorization (MF)) is available to simulate the victim recommender model, UBA conducts simulation experiments with different budgets to attack target users, and then repeats the experiments to assess the treatment effect. In the absence of a reliable surrogate model, we identify a proxy variable for UBA to approximate the heterogeneous treatment effects between target users and items. We have empirically and theoretically validated that the identified proxy variable --- the three-order path number between the target user and item in the user-item interaction graph --- exhibits a strong positive correlation with the recommendation probability of CF models.
Based on the estimated treatment effects, UBA employs a dynamic programming algorithm to compute the optimal budget allocation for all target users, maximizing the overall recommendation probability. 


Since UBA is a model-agnostic framework, we instantiate it on three competitive attack models and conduct experiments on three benchmark datasets. Extensive experiments show the superiority and generalization ability of UBA in various settings, such as different target items, fake user budgets, attack models, and victim models. 
Moreover, we validate the robustness of UBA in the cases of applying defense models although the defense models are effective to some extent. 
To ensure reproducibility, we release our code and data at \url{https://github.com/Wcsa23187/UBA}. 

To summarize, our contributions are threefold.
\begin{itemize}[leftmargin=*]
    \item We highlight the significance of target user attacks 
    and formally inspect the issue of varying attack difficulty on users from a causal perspective. 
    \item We propose the model-agnostic UBA framework, which offers two methods to estimate the heterogeneous treatment effects on target users and calculates the optimal budget allocation to maximize attack performance. 
    \item Extensive experiments reveal the significance of UBA in performing target user attacks across diverse settings. Meanwhile, we validate the robustness of UBA against defense models. 
\end{itemize}

\section{Related Work}
\label{sec:related_work}

In this section, we introduce closely related concepts and literature on uplift modeling and injective attacks. More related studies may refer to Appendix Section~\ref{appendix:related_work}. 

\vspace{3pt}
\noindent\textbf{$\bullet$ Uplift Modeling.}
Uplift, a term commonly used in marketing, usually represents the difference in purchasing actions between customers who receive a promotional offer (the treated group) and those who do not (the control group)~\cite{zhang_unified_2021,liu2022towards}. In causal language, uplift essentially quantifies the causal effect of a treatment (\eg a promotion) on the outcome (\eg purchasing behaviors). Despite extensive research on uplift modeling in the machine learning and marketing communities~\cite{Causal-gutierrez-2017, Ai-LBCF-2022, Tu-Personalized-2021}, the use of uplift modeling in recommendation receives little scrutiny~\cite{xie21causCF,yao21survey,wu2022opportunity}. 
Initial studies only consider the potential of uplift modeling to regulate the exposure proportion of item categories~\cite{yu-MDP2-2022}. 
By contrast, we define the assigned fake user budgets in injective attacks as the treatment and estimate the difference of recommendation probabilities on target users as the uplifts. Based on the estimated uplifts, we aim to determine the best treatment for budget allocation to maximize overall attack performance. 



\vspace{3pt}
\noindent\textbf{$\bullet$ Injective Attacks.}
The objective of injective attacks (\aka shilling attacks) is to promote the recommendations of a target item to all users on the recommender platform~\cite{182952,burke2005limited,huang2021data,zhang2021data,wilson2013power,Nguyen23PGRS}. Given a target item, the attacker optimizes fake user interactions, 
and then the interactions of real users and generated fake users are fed into the victim model for training, improving the recommendation probabilities of the target item to all real users. 

Formally, give a target item $i$ in the item set $\mathcal{I}$, and a set of real users $\mathcal{U}_r$ with their historical interaction matrix $\bm{D}_r\in\{0,1\}^{|\mathcal{U}_r|\times|\mathcal{I}|}$ where $1$ and $0$ indicate users' liked and disliked items, the attacker aims to craft $\bm{D}_f\in\{0,1\}^{|\mathcal{U}_f|\times|\mathcal{I}|}$, the interaction matrix of a set of fake users $\mathcal{U}_f$, for maximizing the attack objective $\mathcal{O}$ on a victim recommender model $\mathcal{M}_\theta$:
\begin{equation}\label{eqn:injective_attack}
\begin{aligned}
        &\max_{\bm{D}_f} \mathcal{O}(\mathcal{M}_{\theta^*}, \mathcal{U}_r, i), \\
        \text{s.t. } &\theta^* = \arg\min_{\theta}\mathcal{L}(\mathcal{M}_{\theta}, \bm{D}); |\mathcal{U}_f|\leq N,
\end{aligned}
\end{equation}
where the victim recommender model $\mathcal{M}_{\theta^*}$ is well trained via the loss function $\mathcal{L}(\cdot)$ calculated on the interactions of both real and fake users, \ie 
$\bm{D}=$$\left[\begin{array}{c}
      \bm{D}_r \\
      \bm{D}_f
\end{array}\right]$. 
Besides, the budget of fake users is limited by a hyper-parameter $N$~\cite{li2016data,zhang2021data,zhang2020practical}. In particular, we detail three key components, attack objective, attack knowledge, and optimization of $\bm{D}_f$, as follows. 

\vspace{3pt}
\textit{\textbf{Attack objective.}}
The attack objective $\mathcal{O}(\mathcal{M}_{\theta^*}, \mathcal{U}_r, i)$ is usually defined as enhancing the probabilities of recommending the target item $i$ to all users $\mathcal{U}_r$ by the victim model $\mathcal{M}_{\theta^*}$. Generally, it can be evaluated by the hit ratio on the real user set $\mathcal{U}_r$, where a user is ``hit'' only when the target item $i$ is ranked into this user's Top-$K$ recommendation list. 

\vspace{3pt}
\textit{\textbf{Attack knowledge.}}
Existing studies 
have different assumptions for 
the knowledge available to the attacker, 
where the knowledge mainly involves the users' interaction data $\bm{D}_r$ and the victim recommender model $\mathcal{M}_{\theta^*}$. Specifically, white-box settings~\cite{fang2018poisoning,fang2020influence,li2016data} might presume both $\bm{D}_r$ and the parameters of $\mathcal{M}_{\theta^*}$ are accessible to the attacker. By contrast, the definitions of gray-box settings vary~\cite{yang2017fake,cao2020adversarial,wu2021fight}. While they consistently assume the accessible $\bm{D}_r$, the usage of $\mathcal{M}_{\theta^*}$ differs~\cite{song2020poisonrec,fan2021attacking,tang2020revisiting,zhang2022targeted,zhang2020practical,wu-Triple-2021}. Some work utilizes the recommendation lists from $\mathcal{M}_{\theta^*}$~\cite{song2020poisonrec,fan2021attacking} while some researchers assume $\mathcal{M}_{\theta^*}$ is totally unavailable and only adopt a surrogate model $\mathcal{S}_{\phi}$ as a replacement for attacking~\cite{tang2020revisiting,zhang2022targeted,zhang2020practical,wu-Triple-2021}. 

\vspace{2pt}
\textit{\textbf{Optimization of $\bm{D}_f$.}}
To adjust $\bm{D}_f$ for maximizing the attack objective, existing methods fall into three lines~\cite{deldjoo2021survey,mobasher2007toward,sharma2013survey}. First, heuristic attackers intuitively increase the co-occurrence of some selected items and the target item $i$ in $\bm{D}_f$ via some heuristic rules, enhancing the popularity of item $i$~\cite{li2022revisiting}. However, such methods cannot directly optimize the attack objectives, leading to poor attack performance~\cite{li2016data,fang2020influence,li2022black}. 
Besides, to maximize the attack objectives, gradient-based methods directly optimize the interactions of fake users~\cite{li2016data,fang2020influence,li2022black} while neural attackers optimize neural networks to generate fake user interactions~\cite{huang2021data,tang2020revisiting,lin2020attacking,Lin_2022}. Their optimization process typically utilizes the recommendations of the victim model or a surrogate model for gradient descent~\cite{pang2018unorganized,xing2013take}. 


\vspace{-0.1cm}
\section{Task Formulation}
\label{sec:task}
In this section, we formulate the task of target user attacks. Besides, we quantify the attack difficulty across users from a causal view. 

\vspace{3pt}
\noindent\textbf{$\bullet$ Target User Attacks.}
Despite the great success of injective attacks, we focus on a novel recommender attack task --- target user attacks, which aim to promote the recommendations of a target item to a group of specific users. This is more reasonable since each target item has its own potential audience. Blindly attacking all users will waste the limited fake user budgets and lead to suboptimal attack performance. Under target user attacks, the attacker can freely specify potential users for attacking based on user features, users' historical interactions, or attack difficulty. 

\vspace{2pt}
\textit{\textbf{Attack objective.}}
Formally, target user attacks change the objective $\mathcal{O}(\mathcal{M}_{\theta^*}, \mathcal{U}_r, i)$ in Eq. (\ref{eqn:injective_attack}) to $\mathcal{O}(\mathcal{M}_{\theta^*}, \mathcal{U}_t, i)$, where $\mathcal{U}_t$ denotes the selected target user group. 

\vspace{2pt}
\textit{\textbf{Attack knowledge.}}
In this work, we adhere to a stricter yet more practical setting: only the interactions of a proportion of real users are accessible, given the fact that the attackers can never collect the interactions of all users. Furthermore, we assume the victim model $\mathcal{M}_{\theta^*}$ is unknown. We consider the situations with (w/) and without (w/o) using a surrogate model $\mathcal{S}_{\phi}$. 

\vspace{2pt}
\textit{\textbf{Optimization of $\bm{D}_f$.}}
Previous methods of injective attacks can be adapted for target user attacks by revising the attack objective to $\mathcal{O}(\mathcal{M}_{\theta^*}, \mathcal{U}_t, i)$. In detail, heuristic attackers~\cite{1167344,burke2005limited,kaur2016shilling,1565730} are able to increase the co-occurrence of target items and the items liked by the target users. Gradient-based and neural attackers may optimize the generation of fake user interactions to maximize $\mathcal{O}(\mathcal{M}_{\theta^*}, \mathcal{U}_t, i)$~\cite{tang2020revisiting,lin2020attacking,li2016data}. For instance, neural attackers~\cite{Lin_2022} can randomly sample a target user as a template for fake user generation to increase the similarity between fake users and the target user, and then forcibly set the fake user feedback on the target item as like, pushing the victim model to recommend the target item to the target user.


\begin{figure*}[t]
\setlength{\abovecaptionskip}{0.1cm}
\setlength{\belowcaptionskip}{-0.3cm}
\centering
\includegraphics[scale=0.68]{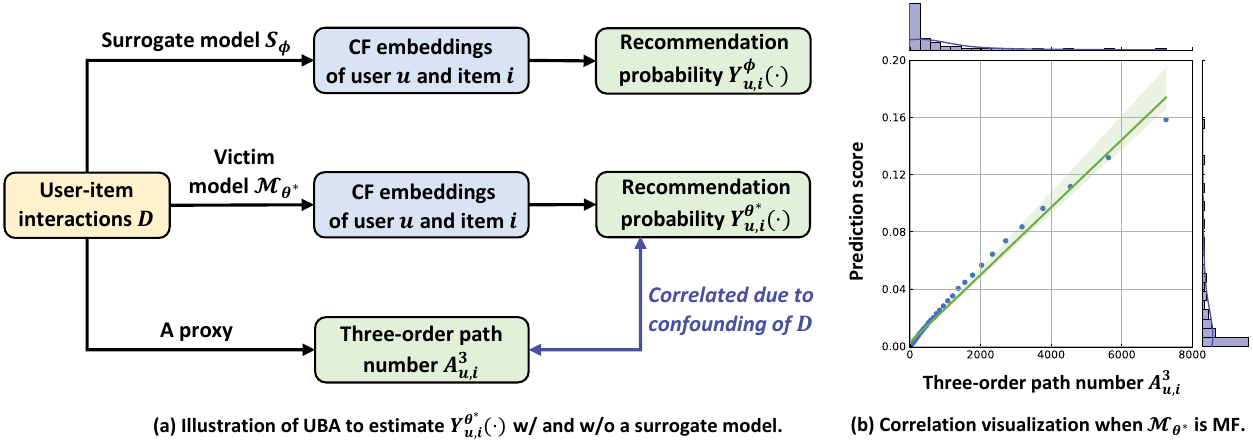}
\caption{Illustration of two estimation methods and the correlation analysis from a causal view, and (b) correlation visualization between three-order path numbers of user-item pairs and their prediction scores by MF, where the correlation coefficients $r=0.9998\approx 1$ and $p=6e^{-86}\ll0.001$ via the Spearman Rank Correlation Test~\cite{spearman} validate the strong correlation (see similar correlation on more CF models in Appendix Section~\ref{appendix:correlation_exp}). }
\label{fig:correlation_vis}
\end{figure*}

\noindent\textbf{$\bullet$ Attack Difficulty.}
However, these intuitive methods ignore the varying attack difficulty on different target users. As illustrated in Figure~\ref{fig:target_users}(b), an increase in the fake user budgets from 4 to 5 might be negligible to uplift the recommendation probabilities on user B while it can significantly enhance that for user A. 
Gradient-based attackers directly optimize fake user interactions for attacking all target users while neural attackers may randomly sample target users as templates to generate fake users by neural networks. 
They are averagely assigning fake user budgets to all target users without considering the varying attack difficulty. 
As a special case, DADA~\cite{yu-MDP2-2022} considers the difficulty of attacking each user regarding a target item, yet it only employs a greedy algorithm to emphasize easier users than difficult users. In this way, DADA fails to quantify the attack difficulty across users and achieve optimal budget allocation to target users. 


Formally, we measure the attack difficulty of a target user $u$ regarding a target item $i$ via a function of recommendation probability \wrt the fake user interactions $\bm{D}_f$, denoted as $Y^{\theta^*}_{u,i}(\bm{D}_f)$, 
which represents the probability of ranking item $i$ into the Top-$K$ recommendation list of user $u$ by the victim model $\mathcal{M}_{\theta^*}$. 
Notably, the fake user interaction matrix $\bm{D}_f$ in $Y^{\theta^*}_{u,i}(\cdot)$ is determined by the attacker with the fake user budget allocation $T$. From a causal view, we formulate the fake user budget allocation on target users as a multi-dimension treatment $T=\{t_u\}_{u\in \mathcal{U}_t}$, where $t_u$ satisfies that $\sum_{u\in \mathcal{U}_t}t_u \leq N$, $t_u \geq 0$, and $t_u \in \mathbb{Z}$. 
Given a target item $i$, a victim model $\mathcal{M}_{\theta^*}$, and fake user interactions $\bm{D}_f$ with treatment $T$, the outcome across all target users is formulated as $\{Y^{\theta^*}_{u,i}(\bm{D}_f(T))\}_{u\in \mathcal{U}_t}$.

Given a target item $i$, a set of target users $\mathcal{U}_t$, and $N$ fake user budgets, the objective of target user attacks can be reformulated to consider attack difficulty as follows: 
\begin{equation}\label{eqn:objective}
\begin{aligned}
        &\max_{T,\bm{D}_f} \mathcal{O}(\mathcal{M}_{\theta^*}, \mathcal{U}_t, i)=\sum_{u\in \mathcal{U}_t}Y^{\theta^*}_{u,i}(\bm{D}_f(T)), \\
        \text{s.t. } & \sum_{u\in \mathcal{U}_t}t_u \leq N; t_u \geq 0, t_u \in \mathbb{Z} \text{ for any } u \in \mathcal{U}_t, 
\end{aligned}
\end{equation}
where we optimize the fake user interaction matrix $\bm{D}_f$ to maximize the overall Top-$K$ recommendation probability on all target users. 
$\bm{D}_f$ is further affected by different attackers\footnote{To keep notation brevity, we omit the notation of attackers in $\bm{D}_f(T)$.} and the budget allocation $T$ across target users, which considers varying attack difficulty. For instance, given a target user $u$ with a budget $t_u$, heuristic attackers can construct $t_u$ fake users with similar interactions to user $u$, increasing the co-occurrence probability of target items and the items liked by user $u$. Gradient-based attackers can optimize the interactions of $t_u$ fake users specifically to maximize the attack objective of target user $u$. Neural attackers might take the target user $u$ as a template to generate the interactions of $t_u$ fake users. Beyond averagely assigning fake user budgets, the attackers can execute target user attacks more purposefully and effectively by considering varying attack difficulty for treatment $T$.

To keep the generality and simplicity of the optimization, we aim to design a model-agnostic solution to first estimate the optimal budget allocation $T^*$ for a given attacker, and then we can apply $T^*$ to the attacker for the final optimization of $\bm{D}_f$.

\vspace{2pt}
\noindent\textbf{$\bullet$ Objective.}
The key to considering the varying attack difficulty for the maximal attack objective lies in 1) estimating $Y^{\theta^*}_{u,i}(\bm{D}_f(t_u))$ of each target user $u$ \wrt different budgets $t_u$~\footnote{Here we change $T$ in $Y^{\theta^*}_{u,i}(\bm{D}_f(\cdot))$ to $t_u$ to denote how $Y^{\theta^*}_{u,i}(\cdot)$ of target user $u$ changes \wrt the user $u$' own budget $t_u$. The mutual interference of fake user budgets among target users is tough to measure, which is left for further exploration.} (see the curves in Figure~\ref{fig:target_users}(b)), and 2) calculating the optimal treatment $T^*$ to allocate fake users and generate $\bm{D}_f$ for attacking.

\section{Uplift-guided Budget Allocation}
\label{sec:method}

We propose a UBA framework with two methods to estimate the treatment effects $Y^{\theta^*}_{u,i}(\bm{D}_f(t_u))$ on each target user, and then calculate the optimal treatment $T^*$ for target user attacks. Lastly, we detail how to instantiate UBA on existing attackers.

\subsection{Treatment Effect Estimation}\label{sec:treatment_estimation}
Since the victim model $\mathcal{M}_{\theta^*}$ in Eq. (\ref{eqn:objective}) is unknown to attackers, we propose two methods w/ and w/o the surrogate model $\mathcal{S}_{\phi}$ to estimate $Y^{\theta^*}_{u,i}(\bm{D}_f(t_u))$ \wrt varying budgets $t_u$. 

\vspace{3pt}
\noindent\textbf{$\bullet$ Estimation via Simulation Experiments.}
If $\mathcal{S}_{\phi}$ is available to replace $\mathcal{M}_{\theta^*}$, UBA can utilize $\mathcal{S}_{\phi}$ to do simulation experiments for the estimation of $Y^{\theta^*}_{u,i}(\bm{D}_f(t_u))$ with varying $t_u$. 
In detail, we repeat $E$ times of simulation experiments with different random seeds. 
In each experiment, given a target item $i$, an attacker, and a surrogate model $\mathcal{S}_\phi$, we do the following steps: 1) assigning the same $t_u$ fake users to all target users; 2) using the attacker to generate the fake user matrix $\bm{D}_f$ based on the assigned budget; and 3) leveraging $\bm{D}_f$ to attack $\mathcal{S}_\phi$ and obtain its Top-$K$ recommendations for each target user. After $E$ times of experiments with different random seeds for training ($E\approx10$ in our implementation), we can approximate $Y^{\theta^*}_{u,i}(\bm{D}_f(t_u))$ via the hit ratio of target user $u$: 
\begin{equation}\label{eqn:RCT_Y}
\begin{aligned}
Y^{\theta^*}_{u,i}(\bm{D}_f(t_u))\approx \frac{E'_{t_u}}{E},
\end{aligned}
\end{equation}
where $E'_{t_u}$ is the number of successfully promoting the target item $i$ into the Top-$K$ recommendations of target user $u$ by using $t_u$ budgets in $E$ times of experiments. 
Similarly, we can vary $t_u \in \{1,2,...,H\}$ with $H\ll N$ to estimate $Y^{\theta^*}_{u,i}(\bm{D}_f(t_u))$ with different fake user budgets\footnote{The potential budget of each target user is far less than the total budget number $N$ because $N$ budgets are assigning to a group of target users.}.

\vspace{3pt}
\noindent\textbf{$\bullet$ Estimation via High-order Interaction Path.}
Although UBA can estimate the treatment effects via simulation experiments, they rely on a reliable surrogate model and require repeated simulation experiments. To shake off these shackles, we propose another treatment effect estimation method w/o using surrogate models and simulation experiments. 

We analyze the reasons for the varying attack difficulty across target users. 
Recommender models, including $\mathcal{M}_{\theta^*}$ and $\mathcal{S}_\phi$, learn CF embeddings of users and items from interactions as illustrated in Figure~\ref{fig:correlation_vis}(a), and then measure the similarity between their CF embeddings to output 
prediction scores for item ranking. 
The varying attack difficulty is attributed to the distinct CF embeddings, and essentially stems from the different similarities \wrt interactions. 
To eliminate the dependence on surrogate models, we consider finding a proxy of interactions for the estimation of $Y^{\theta^*}_{u,i}(\bm{D}_f(t_u))$. 
We inspect the number of high-order paths between a target user and a target item in the user-item interaction graph, discovering that it has a strong positive correlation with 
the user-item prediction scores as shown in Figure~\ref{fig:correlation_vis}(b). 
Formally, if we define $\bm{A}=\left[\begin{array}{cc}
      \bm{0} & \bm{D}_r \\
      \bm{D}^T_r & \bm{0}
\end{array}\right]$, the three-order path number is the value in $\bm{A}^3$ indexed by a user and an item, which describes the easiness of connecting this user-item pair via some intermediate users and items (see more explanation in Appendix Section~\ref{appendix:correlation_exp}). 

\begin{proposition}\label{pro:A3_path}
Given a user and an item without historical interactions, their 
prediction score
by a CF model is positively correlated with their three-order\footnote{We find that the correlation exists with multiple different order numbers (see Appendix Section~\ref{appendix:correlation_exp}). To keep simplicity, we select the smallest order for investigation.} path number in $\bm{A}^3$. 
\end{proposition}
We theoretically analyze the rationality and robustness of Proposition~\ref{pro:A3_path} \wrt different CF models in Appendix Section~\ref{appendix:P_1_proofs}. In Appendix Section~\ref{appendix:correlation_exp}, we also present extensive experiments to validate the wide existence of such a correlation~\cite{Paudel16Up}. 

Intuitively, CF models utilize the interaction similarity across users and items for recommendations, and three-order path numbers describe the closeness between users and items in the interaction graph~\cite{he2020lightgcn}. The user-item pairs with larger three-order path numbers are intuitively closer to each other in the CF embedding space, leading to higher prediction scores (see Figure~\ref{fig:correlation_vis}(b)). 
Since the prediction scores are used to rank Top-$K$ recommendations, the three-order path number is also positively correlated with the Top-$K$ recommendation probability $Y^{\theta^*}_{u,i}(\cdot)$.   
Moreover, from a causal view in Figure~\ref{fig:correlation_vis}(a), user-item interactions act as a confounder, causing the correlation between three-order path numbers and Top-$K$ recommendation probabilities. 

Thanks to such a positive correlation, we can approximate $Y^{\theta^*}_{u,i}(\cdot)$ by the weighted three-order path number between user $u$ and item $i$, denoted as $\alpha\cdot(\bm{A}^3_{u,i})^\beta$, where $\alpha$ and $\beta$ are two hyper-parameters to adapt for the correlation. Thereafter, to estimate $Y^{\theta^*}_{u,i}(\bm{D}_f(t_u))$ with varying budgets $t_u$, we only need to assess $\alpha\cdot(\bm{A}^3_{u,i})^\beta$ after injecting $t_u$ fake users to maximize the three-order path number between $u$ and $i$. By analyzing the relationships between fake user interactions and the three-order path number, we have the following finding. 

\begin{proposition}\label{pro:A3_path_user}
The three-order path number between target user $u$ and target item $i$ ($\bm{A}^3_{u,i}$) is equivalent to the weighted sum of the intermediate users who liked item $i$ in $\bm{A}$, where the weights are their interaction similarities with target user $u$, \ie the number of mutually liked items. 
\end{proposition}
Please refer to Appendix Section~\ref{appendix:P_2_proofs} for the proofs of Proposition~\ref{pro:A3_path_user}. In the light of Proposition~\ref{pro:A3_path_user}, given a target user $u$ and a target item $i$, we can construct $t_u$ fake users with the largest interaction similarities with target user $u$ and a like interaction with target item $i$. 
Consequently, we can obtain the optimal three-order path number in $(\bm{A}')^3$ to estimate $Y^{\theta^*}_{u,i}(\bm{D}_f(t_u))$ via $Y^{\theta^*}_{u,i}(\bm{D}_f(t_u)) \approx \alpha\cdot((\bm{A}')^3_{u,i})^\beta$, where $\bm{A'}=\left[\begin{array}{ccc}
      \bm{0} & \bm{0} & \bm{D}_r \\
      \bm{0} & \bm{0} & \bm{D}_f \\
      \bm{D}^T_r & \bm{D}^T_f & \bm{0}
\end{array}\right]$, \ie the symmetric interaction matrix with both real and fake users.

\subsection{Budget-constrained Treatment Optimization}
After estimating the treatment effect $Y^{\theta^*}_{u,i}(\bm{D}_f(t_u))$, the selection of treatment $T$ in Eq. (\ref{eqn:objective}) becomes a budget-constrained optimization problem~\cite{Ai-LBCF-2022}. To calculate the optimal treatment $T^*$, we implement a dynamic programming algorithm based on the idea of the knapsack problem (see Appendix Section~\ref{appendix:DP} for details). 
Afterward, we can allocate optimal budgets to each target user to enhance existing attackers for superior overall attack performance. 

\vspace{3pt}
\noindent\textbf{$\bullet$ Instantiation on Existing Attackers.}
Given the optimal $T^*$, existing attackers, including heuristic, gradient-based, and neural attackers, can allocate fake users accordingly and utilize their own strategies to construct fake user interactions (see analysis of Eq. (\ref{eqn:objective}) in Section~\ref{sec:task} for details). These fake users are subsequently used to attack victim recommender models.

\vspace{3pt}
\noindent\textbf{$\bullet$ UBA for Enhancing Recommendation Security.} A superior and transparent attacker can always inspire stronger defense methods. Recommender platforms can utilize UBA to improve their defense models through adversarial training, thereby enhancing the security of recommendations.

\section{Experiments}
\label{sec:exp}

In this section, we conduct extensive experiments to answer the following research questions:

\textbf{$-$ RQ1:} How does UBA enhance existing attackers in the task of target user attacks?

\textbf{$-$ RQ2:} How does UBA generalize across various settings (\eg different victim models, budgets, and accessible interactions)?

\textbf{$-$ RQ3:} How do UBA and other attackers perform if defense models are applied?

\begin{table*}[t]
\setlength{\abovecaptionskip}{0cm}
\setlength{\belowcaptionskip}{0cm}
\caption{Evaluation on three real-world datasets when overall fake user budget is 100. We show the results of HR@10, NDCG@10, and MRR@10. The best results for each backend model are bold and the second-best ones are underlined. $*$ implies the improvements over the best baseline ``Target'' are statistically significant ($p$-value<0.05) under $t$-test.}
\label{tab:main}
\begin{center}
\setlength{\tabcolsep}{0.66mm}{
\resizebox{\textwidth}{!}{
\begin{tabular}{lcccccccccccccccccc}
\hline
 & \multicolumn{6}{c}{\textbf{ML-1M}} & \multicolumn{6}{c}{\textbf{Yelp}} & \multicolumn{6}{c}{\textbf{Amazon}} \\
 & \multicolumn{3}{c}{\textbf{Popular   item}} & \multicolumn{3}{c}{\textbf{Unpopular   item}} & \multicolumn{3}{c}{\textbf{Popular   item}} & \multicolumn{3}{c}{\textbf{Unpopular   item}} & \multicolumn{3}{c}{\textbf{Popular   item}} & \multicolumn{3}{c}{\textbf{Unpopular   item}} \\
 & \textbf{HR} & \textbf{NDCG} & \textbf{MRR} & \textbf{HR} & \textbf{NDCG} & \textbf{MRR} & \textbf{HR} & \textbf{NDCG} & \textbf{MRR} & \textbf{HR} & \textbf{NDCG} & \textbf{MRR} & \textbf{HR} & \textbf{NDCG} & \textbf{MRR} & \textbf{HR} & \textbf{NDCG} & \multicolumn{1}{l}{\textbf{MRR}} \\ \hline
\textbf{Beofore Attack} & 0.0000 & 0.0000 & 0.0000 & 0.0000 & 0.0000 & 0.0000 & 0.0000 & 0.0000 & 0.0000 & 0.0000 & 0.0000 & 0.0000 & 0.0200 & 0.0035 & 0.0020 & 0.0000 & 0.0000 & 0.0000 \\
\textbf{Random Attack} & 0.0000 & 0.0000 & 0.0000 & 0.0000 & 0.0000 & 0.0000 & 0.0200 & 0.0028 & 0.0022 & 0.0200 & 0.0028 & 0.0022 & 0.2000 & 0.0235 & 0.0199 & 0.2200 & 0.0262 & 0.0220 \\
\textbf{Segment Attack} & 0.0200 & 0.0040 & 0.0022 & 0.0200 & 0.0035 & 0.0035 & 0.0000 & 0.0000 & 0.0000 & 0.0400 & 0.0039 & 0.0040 & 0.1600 & 0.0195 & 0.0160 & 0.1400 & 0.0180 & 0.0155 \\
\textbf{Bandwagon Attack} & 0.0200 & 0.0035 & 0.0020 & 0.0000 & 0.0000 & 0.0000 & 0.0200 & 0.0029 & 0.0025 & 0.0200 & 0.0027 & 0.0020 & 0.2000 & 0.0245 & 0.0222 & 0.1600 & 0.0204 & 0.0177 \\
\textbf{Average Attack} & 0.0000 & 0.0000 & 0.0000 & 0.0000 & 0.0000 & 0.0000 & 0.0200 & 0.0031 & 0.0029 & 0.0200 & 0.0029 & 0.0025 & 0.1800 & 0.0217 & 0.0200 & 0.1400 & 0.0180 & 0.0355 \\
\textbf{WGAN} & 0.0200 & 0.0039 & 0.0025 & 0.0000 & 0.0000 & 0.0000 & 0.0200 & 0.0029 & 0.0025 & 0.0200 & 0.0027 & 0.0020 & 0.1400 & 0.0180 & 0.0155 & 0.1800 & 0.0227 & 0.0225 \\
\textbf{DADA-DICT} & 0.0400 & 0.0071 & 0.0073 & 0.0800 & 0.0104 & 0.0089 & 0.0200 & 0.0036 & 0.0040 & 0.1200 & 0.0169 & 0.0150 & 0.4200 & 0.0517 & 0.0420 & 0.3200 & 0.0448 & 0.0400 \\
\textbf{DADA-DIV} & 0.0000 & 0.0000 & 0.0000 & 0.0400 & 0.0057 & 0.0044 & 0.0000 & 0.0000 & 0.0000 & 0.1400 & 0.0220 & 0.0199 & 0.2600 & 0.0333 & 0.0260 & 0.3200 & 0.0473 & 0.0457 \\
\textbf{DADA} & 0.0600 & 0.0112 & 0.0123 & 0.1200 & 0.0156 & 0.0171 & 0.0400 & 0.0053 & 0.0080 & 0.1600 & 0.0268 & 0.0266 & 0.5400 & 0.0667 & 0.0540 & 0.4400 & 0.0576 & 0.0488 \\ \hline
\textbf{AIA} & 0.0200 & 0.0061 & 0.0033 & 0.0200 & 0.0127 & 0.0111 & 0.0000 & 0.0000 & 0.0000 & 0.0600 & 0.0054 & 0.0032 & 0.2800 & 0.0375 & 0.0311 & 0.2800 & 0.0393 & 0.0350 \\
\textbf{\quad+Target} & 0.2000 & 0.0244 & 0.0222 & 0.3800 & 0.0428 & 0.0356 & 0.0000 & 0.0000 & 0.0000 & 0.0600 & 0.0082 & 0.0075 & 0.5800 & 0.0780 & 0.0725 & 0.6400 & 0.0846 & 0.0711 \\
\textbf{\quad+ UBA(w/o $\mathcal{S}_{\phi}$)} & {\ul 0.2400} & {\ul 0.0315} & {\ul 0.0300} & {\ul 0.4400} & {\ul 0.0449} & {\ul 0.0400} & {\ul 0.0200} & {\ul 0.0029} & {\ul 0.0025} & {\ul 0.1200} & {\ul 0.0191} & {\ul 0.0269} & {\ul 0.6600} & {\ul 0.0960} & {\ul 0.0942} & {\ul 0.7400} & {\ul 0.1007} & {\ul 0.0925} \\
\textbf{\quad+ UBA(w/ $\mathcal{S}_{\phi}$)} & \textbf{0.2600*} & \textbf{0.0431*} & \textbf{0.0520*} & \textbf{0.5800*} & \textbf{0.0556*} & \textbf{0.0514*} & \textbf{0.1000*} & \textbf{0.0178*} & \textbf{0.0250*} & \textbf{0.1400*} & \textbf{0.0255*} & \textbf{0.0279*} & \textbf{0.7200*} & \textbf{0.1115*} & \textbf{0.1200*} & \textbf{0.8200*} & \textbf{0.1162*} & \textbf{0.1171*} \\ \hline
\textbf{AUSH} & 0.0200 & 0.0047 & 0.0050 & 0.0000 & 0.0000 & 0.0000 & 0.0000 & 0.0000 & 0.0000 & 0.0000 & 0.0000 & 0.0000 & 0.2600 & 0.0347 & 0.0288 & 0.3200 & 0.0448 & 0.0400 \\
\textbf{\quad+Target} & 0.2400 & 0.0349 & 0.0340 & 0.3600 & 0.0531 & 0.0422 & 0.0800 & 0.0111 & 0.0100 & 0.1000 & 0.0119 & 0.0100 & 0.6400 & 0.0887 & 0.0800 & 0.7400 & 0.1064 & 0.1057 \\
\textbf{\quad+ UBA(w/o $\mathcal{S}_{\phi}$)} & {\ul 0.2800} & {\ul 0.0380} & {\ul 0.0467} & {\ul 0.4200} & {\ul 0.0646} & {\ul 0.0550} & {\ul 0.1000} & {\ul 0.0138} & {\ul 0.0142} & {\ul 0.0800} & {\ul 0.0137} & {\ul 0.0160} & {\ul 0.7400} & {\ul 0.1064} & {\ul 0.1057} & {\ul 0.7400} & {\ul 0.1137} & {\ul 0.1233} \\
\textbf{\quad+ UBA(w/ $\mathcal{S}_{\phi}$)} & \textbf{0.3200*} & \textbf{0.0474*} & \textbf{0.0533*} & \textbf{0.5200*} & \textbf{0.0902*} & \textbf{0.0828*} & \textbf{0.1400*} & \textbf{0.0255*} & \textbf{0.0279*} & \textbf{0.1600*} & \textbf{0.0376*} & \textbf{0.0533*} & \textbf{0.8600*} & \textbf{0.1299*} & \textbf{0.1433*} & \textbf{0.8000*} & \textbf{0.1142*} & \textbf{0.1142*} \\ \hline
\textbf{Legup} & 0.0400 & 0.0062 & 0.0054 & 0.1000 & 0.0024 & 0.0021 & 0.0600 & 0.0078 & 0.0067 & 0.0400 & 0.0043 & 0.0050 & 0.0400 & 0.0063 & 0.0050 & 0.2000 & 0.0235 & 0.0199 \\
\textbf{\quad+Target} & 0.1400 & 0.0187 & 0.0200 & 0.3200 & 0.0539 & 0.0412 & 0.1600 & 0.0251 & 0.0229 & 0.1800 & 0.0302 & 0.0300 & 0.6000 & 0.0742 & 0.0600 & 0.5000 & 0.0676 & 0.0625 \\
\textbf{\quad+ UBA(w/o $\mathcal{S}_{\phi}$)} & {\ul 0.1800} & {\ul 0.0239} & {\ul 0.0225} & {\ul 0.3200} & {\ul 0.0604} & {\ul 0.0525} & {\ul 0.2000} & {\ul 0.0283} & {\ul 0.0222} & {\ul 0.2600} & {\ul 0.0411} & {\ul 0.0433} & {\ul 0.7400} & {\ul 0.0961} & {\ul 0.0822} & {\ul 0.6800} & {\ul 0.0887} & {\ul 0.0755} \\
\textbf{\quad + UBA(w/ $\mathcal{S}_{\phi}$)} & \textbf{0.2600*} & \textbf{0.0480*} & \textbf{0.0371*} & \textbf{0.3600*} & \textbf{0.0825*} & \textbf{0.0743*} & \textbf{0.2400*} & \textbf{0.0365*} & \textbf{0.0342*} & \textbf{0.2800*} & \textbf{0.0411*} & \textbf{0.0466*} & \textbf{0.8600*} & \textbf{0.1216*} & \textbf{0.1228*} & \textbf{0.7600*} & \textbf{0.1086*} & \textbf{0.1085*} \\ \hline
\end{tabular}
}}
\end{center}
\end{table*}

\vspace{3pt}
\noindent\textbf{$\bullet$ Datasets and Metric.} We evaluate the attackers on the three real-world datasets: MovieLens-1M (ML-1M)~\cite{10.1145/2827872}, Amazon-Game (Amazon)~\cite{10.1145/3018661.3018696}, and Yelp~\cite{asghar2016yelp}. Table~\ref{tab:datasets} displays the statistics of the datasets. In addition to the commonly used ML-1M and Amazon datasets in previous recommender attack work~\cite{lin2020attacking,li2022revisiting}, we also select Yelp for experiments due to its larger scale and sparsity. In this way, we can analyze target user attacks in more diverse scenarios. We treat target items as the positive items and use a widely used metric Hit Ratio@$K$ (HR@$K$)~\cite{li2022revisiting} to measure how many target users receive target item recommendation, where $K=10$ or $20$, denoting the length of recommendation lists. Additionally, we introduce NDCG@$K$ and MRR@$K$ to consider the ranking positions of target items in the recommendation lists. 

\begin{table}[t]
\renewcommand\arraystretch{0.9}
\setlength{\abovecaptionskip}{0cm}
\setlength{\belowcaptionskip}{0cm}
\caption{Statistics of the three datasets.}
\label{tab:datasets}
\begin{center}
\setlength{\tabcolsep}{2mm}{
\resizebox{0.48\textwidth}{!}{
\begin{tabular}{cccccc}
\toprule
\multicolumn{1}{l}{} & ${|\mathcal{U}_r|}$  & ${|\mathcal{I}|}$  & \textbf{\#Interactions} & \textbf{Sparity} & \textbf{K-core} \\ \midrule
\textbf{ML-1M} & { 5950} & { 3702} & 567,533 & 0.257\% & 10 \\ 
\textbf{Amazon} & { 3179} & { 5600} & 38, 596 & 0.216\% & 10 \\ 
\textbf{Yelp} & { 54632} & { 34474} & 1,334,942 & 0.070\% & 10 \\ \bottomrule
\end{tabular}
}
}
\end{center}
\vspace{-0.4cm}
\end{table}

\vspace{3pt}
\noindent\textbf{$\bullet$ Data Processing.} For data processing, we conduct a 10-core filtering on all three datasets to ensure data quality. 
Besides, all three datasets consist of explicit user feedback such as ratings, which might be required by some existing attackers. We follow the default requirements of the attackers to provide explicit or implicit feedback. 
For the victim recommender models and surrogate recommender models, we employ the common implicit feedback with binary values $\{0,1\}$ in recommender training. Following previous studies~\cite{he2020lightgcn}, we map historical interactions greater than 3 to likes with label 1 and the remaining to dislikes with label 0.

\vspace{3pt}
\noindent\textbf{$\bullet$ Baselines.} 
We compare with the following baselines: heuristic attackers including Random~\cite{kaur2016shilling}, Segment~\cite{1565730}, Bandwagon~\cite{1565730}, and Average Attacks~\cite{lam2004shilling}, the state-of-the-art {gradient-based} attacker DADA with its variants~\cite{li2022revisiting}, and neural attackers including WGAN~\cite{https://doi.org/10.48550/arxiv.1701.07875}, AIA~\cite{tang2020revisiting}, AUSH~\cite{lin2020attacking}, and Leg-UP~\cite{Lin_2022}. 
To adapt to target user attacks, we propose a model-agnostic baseline named ``Target'' by changing the attack objective from $\mathcal{O}(\mathcal{M}_{\theta^*}, \mathcal{U}_r, i)$ to $\mathcal{O}(\mathcal{M}_{\theta^*}, \mathcal{U}_t, i)$ as discussed in Section~\ref{sec:task}. 
We implement ``Target'' and our UBA framework to three competitive backend attackers for comparison. 
We move the hyper-parameter tuning to Appendix Sections~\ref{appendix:hyper-parameter}.  

\vspace{3pt}
\noindent\textbf{$\bullet$ Victim and Surrogate Models.} 
Following~\cite{li2022revisiting}, we choose representative MF~\cite{koren2009matrix}, NCF~\cite{He2017Neural}, and LightGCN~\cite{he2020lightgcn} as victim models and utilize the simplest MF as the surrogate model. Table~\ref{tab:main} shows the results of LightGCN as the victim model due to space limitation. The robustness of attacking MF and NCF is analyzed in Figure~\ref{fig:generation}.


\vspace{3pt}
\noindent\textbf{$\bullet$ Selection of Target Items.} To verify the robustness of UBA on different target items, we test UBA on popular items and unpopular items, respectively. This is because the attack performance on popular and unpopular items might vary. We divide all items into five groups according to their popularity, \ie historical interaction numbers, and then we randomly select the popular item and the unpopular item as the target item from the most popular group and the most unpopular group, respectively. For each target item, we run five attack processes by changing random seeds and report the average performance. 

\vspace{3pt}
\noindent\textbf{$\bullet$ Selection of Target Users.} 
Notably, target users can be specified by user IDs, attributes (\eg gender), and interactions. 
In this work, we try to find out the potential target users who might be interested in the target item. To achieve this, we select all users who have interacted with the target item category. For example, given an action movie as the target item, we find out all the users who have interacted with at least one action movie. 
Thereafter, we rank all the selected users via their interaction number with the target item category. 
Only the users with an interaction number with the target item category smaller than a threshold (10) are further selected, from which we randomly sample some users who are hard to attack into the target user group. In this way, we can select some target users who are possibly interested in this item category yet hard to attack. 
We usually select 50 target users in our experiments while we analyze the effect of different target user numbers in Table~\ref{tab:target_user}.

\subsection{Target User Attack Performance}

\textbf{$\bullet$ Overall Comparison (RQ1).} 
In Table~\ref{tab:main}, we report the attack performance on target users. As an extension, we present the attack results of all users in Appendix Section~\ref{appendix:overall_all_users}. 
From Table~\ref{tab:main}, we have the following findings. 
\begin{itemize}[leftmargin=*]
    \item [1)] \textbf{Superiority of DADA and ``Target''.} Most attackers (heuristic and neural ones) cannot achieve satisfactory performance on target user attacks. By contrast, DADA with variants and ``Target'' exhibit superior results, revealing the importance of distinguishing users for attacking. DADA and DADA-DICT utilize a greedy algorithm to allocate more budgets to easy users while ``Target'' concentrates fake user budgets to attack target users. 
    
    \item [2)] \textbf{Effectiveness of UBA.} Both UBA w/ and w/o $\mathcal{S}_\phi$ significantly enhance the attack performance of three backend attackers (AIA, AUSH, and Leg-UP) by a large margin on three datasets. In addition, UBA also surpasses ``Target'', further validating the superiority of considering the varying attack difficulty. Due to estimating the attack difficulty across target users, UBA can rationally allocate the fake user budgets to maximize the overall recommendation probabilities. The users with large uplifts of recommendation probabilities will be favored, thus enhancing the attack performance.

    \item [3)] \textbf{UBA(w/ $\mathcal{S}_\phi$) outperforms UBA(w/o $\mathcal{S}_\phi$).} This is reasonable, as the surrogate model $\mathcal{S}_\phi$ can assist in accurately estimating the attack difficulty through simulation experiments. $\mathcal{S}_\phi$ may serve as a reliable substitute for the victim model $\mathcal{M}_{\theta^*}$ in such estimation because they commonly leverage CF information for recommendations. Despite the better performance of UBA(w/ $\mathcal{S}_\phi$), UBA(w/o $\mathcal{S}_\phi$) does not need the simulation experiments, reducing the computation costs. As such, UBA(w/o $\mathcal{S}_\phi$) is also a favorable approach that balances effectiveness and efficiency. 
\end{itemize}

\vspace{3pt}

\begin{figure*}[t]
\setlength{\abovecaptionskip}{0.05cm}
\setlength{\belowcaptionskip}{-0.30cm}
\centering
\includegraphics[scale=0.44]{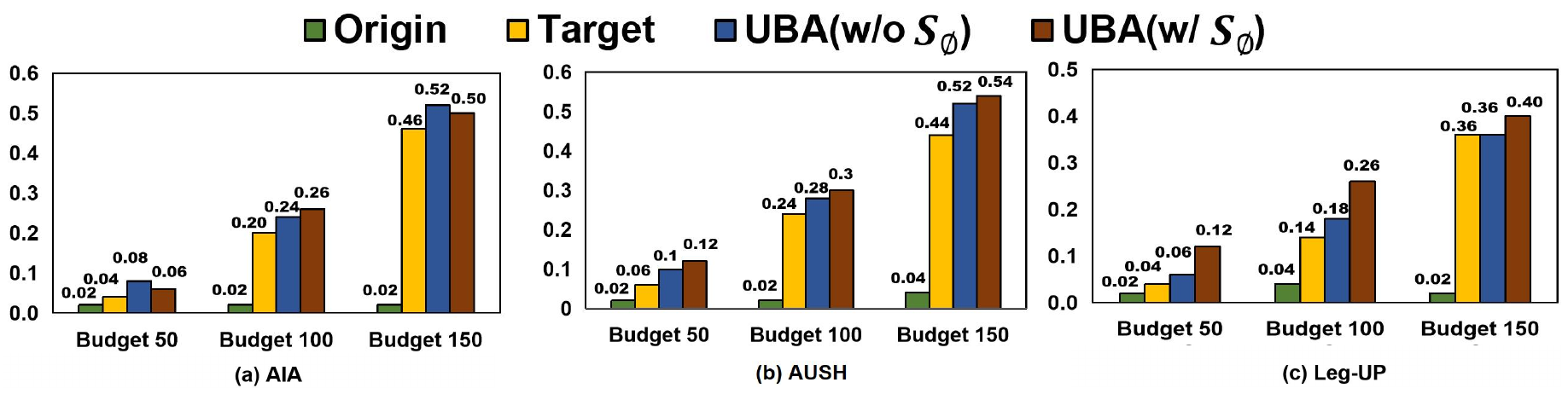}
\caption{Performance comparison \wrt HR@10 under different attack budgets.}
\label{fig:budget}
\end{figure*}

\noindent\textbf{$\bullet$ Varying Fake User Budgets (RQ2).} Figure~\ref{fig:budget} depicts the attack performance on ML-1M with varying budgets. The results on Amazon and Yelp with similar trends are omitted to save space. From the figure, we can find that: 
1) under different budgets, UBA w/ and w/o $\mathcal{S}_{\phi}$ usually achieve better attack performance than the original method and ``Target'' on three backend attackers. This verifies the robustness of UBA \wrt fake user budgets. 
And 2) the relative improvements from ``Target'' to UBA is significantly large when the budget is small such as budget$=50$. Such observation on three attackers demonstrates that UBA holds high practical value in real-world recommender attack scenarios, given that the number of fake users an attacker can manage is often very limited~\cite{li2022revisiting}.

\begin{figure}[t]
\setlength{\abovecaptionskip}{0cm}
\setlength{\belowcaptionskip}{-0.30cm}
  \centering 
  \subfigure{
    \includegraphics[scale=0.455]{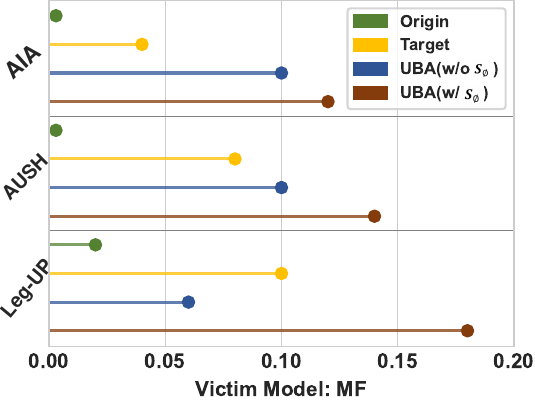}} 
  \subfigure{
    \includegraphics[scale=0.455]{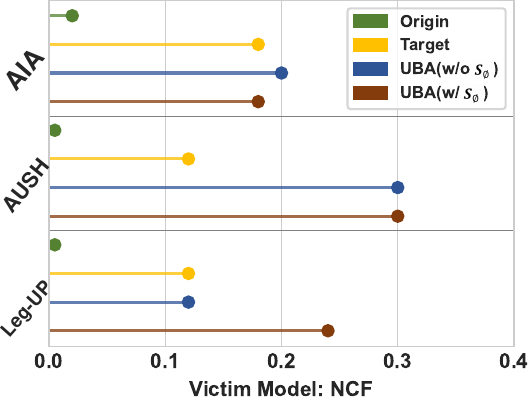}} 
  \caption{Generalization of UBA \wrt HR@10 across different victim models.}
  \label{fig:generation}
  \vspace{0cm}
\end{figure}

\vspace{3pt}
\noindent\textbf{$\bullet$ Different Victim Models (RQ2).} 
Figure~\ref{fig:generation} visualizes the attack results of using MF and NCF as victim models on ML-1M. By inspecting Figure~\ref{fig:generation} and the results of LightGCN in Table~\ref{tab:main}, we can have the following observations. 
1) UBA shows better performance than ``Origin'' and ``Target'' in most cases, indicating good generalization ability of UBA across different victim models. Indeed, the generalization of UBA lies in the accurate estimation of the attack difficulty on target users, \ie the heterogeneous treatment effects. By exploiting the CF information with and without using surrogate models, our proposed UBA is theoretically applicable to most CF victim models. 
And 2) the hit ratios on MF are significantly smaller than those on NCF and LigntGCN. This phenomenon commonly exists across different backend attackers and attack strategies. We attribute the possible reason to that NCF and LightGCN utilize more advanced neural networks to fully exploit CF information in user-item interactions, resulting in a more severe fitting of fake user interactions. Consequently, the fake users lead to a higher attack performance on NCF and LightGCN. This finding is critical, inspiring us to consider the security issues when devising advanced recommender models.

\begin{table}[t]
\setlength{\abovecaptionskip}{0cm}
\setlength{\belowcaptionskip}{0cm}
\caption{Performance with different proportions of user interactions accessible to attackers.}
\label{tab:data_ratio}
\begin{center}
\setlength{\tabcolsep}{0.3mm}{
\resizebox{0.48\textwidth}{!}{
\begin{tabular}{lcccccccc}
\toprule
\textbf{Data Ratio} & \multicolumn{2}{c}{\textbf{10\%}} & \multicolumn{2}{c}{\textbf{20\%}} & \multicolumn{2}{c}{\textbf{50\%}} & \multicolumn{2}{c}{\textbf{80\%}} \\
\multicolumn{1}{l}{} & \textbf{HR@10} & \textbf{HR@20} & \textbf{HR@10} & \textbf{HR@20} & \textbf{HR@10} & \textbf{HR@20} & \textbf{HR@10} & \textbf{HR@20} \\ \midrule
\multicolumn{1}{l}{\textbf{Before Attack}} & 0.02 & 0.02 & 0.00 & 0.02 & 0.00 & 0.02 & 0.00 & 0.02 \\
\multicolumn{1}{l}{\textbf{Random Attack}} & 0.00 & 0.02 & 0.00 & 0.00 & 0.00 & 0.02 & 0.02 & 0.02 \\
\multicolumn{1}{l}{\textbf{Segment Attack}} & 0.00 & 0.02 & 0.02 & 0.02 & 0.02 & 0.02 & 0.02 & 0.02 \\
\multicolumn{1}{l}{\textbf{Bandwagon Attack}} & 0.02 & 0.02 & 0.02 & 0.02 & 0.02 & 0.02 & 0.00 & 0.04 \\
\multicolumn{1}{l}{\textbf{Average Attack}} & 0.00 & 0.00 & 0.00 & 0.00 & 0.00 & 0.02 & 0.00 & 0.00 \\
\multicolumn{1}{l}{\textbf{WGAN}} & 0.00 & 0.02 & 0.02 & 0.02 & 0.00 & 0.02 & 0.00 & 0.02 \\
\multicolumn{1}{l}{\textbf{DADA-DICT}} & 0.02 & 0.04 & 0.04 & 0.04 & 0.00 & 0.04 & 0.04 & 0.04 \\ 
\multicolumn{1}{l}{\textbf{DADA-DIV}} & 0.02 & 0.02 & 0.00 & 0.04 & 0.02 & 0.04 & 0.02 & 0.08 \\
\multicolumn{1}{l}{\textbf{DADA}} & 0.04 & 0.04 & 0.06 & 0.10 & 0.06 & 0.12 & 0.08 & 0.12 \\\midrule
\multicolumn{1}{l}{\textbf{AIA}} & 0.00 & 0.02 & 0.02 & 0.02 & 0.02 & 0.02 & 0.02 & 0.06 \\
\multicolumn{1}{l}{\textbf{\quad +Target}} & 0.12 & 0.28 & 0.20 & 0.38 & 0.22 & 0.36 & 0.20 & 0.42 \\
\multicolumn{1}{l}{\textbf{\quad + UBA(w/o $\mathcal{S}_\phi$)}} & {\ul 0.22} & {\ul 0.36} & {\ul 0.24} & {\ul 0.40} & {\ul 0.22} & {\ul 0.42} & {\ul 0.28} & {\ul 0.44} \\
\multicolumn{1}{l}{\textbf{\quad + UBA(w/ $\mathcal{S}_\phi$)}} & \textbf{0.22} & \textbf{0.42} & \textbf{0.26} & \textbf{0.44} & \textbf{0.30} & \textbf{0.48} & \textbf{0.34} & \textbf{0.48} \\ \midrule
\multicolumn{1}{l}{\textbf{AUSH}} & 0.00 & 0.04 & 0.02 & 0.02 & 0.04 & 0.06 & 0.04 & 0.08 \\
\multicolumn{1}{l}{\textbf{\quad +Target}} & 0.16 & 0.34 & 0.24 & 0.36 & 0.24 & 0.36 & 0.30 & 0.40 \\
\multicolumn{1}{l}{\textbf{\quad + UBA(w/o $\mathcal{S}_\phi$)}} & {\ul 0.20} & {\ul 0.42} & {\ul 0.28} & {\ul 0.40} & {\ul 0.26} & {\ul 0.42} & \textbf{ 0.32} & {\ul 0.44} \\
\multicolumn{1}{l}{\textbf{\quad + UBA(w/ $\mathcal{S}_\phi$)}} & \textbf{0.28} & \textbf{0.44} & \textbf{0.32} & \textbf{0.42} & \textbf{0.30} & \textbf{0.46} & {\ul 0.28} & \textbf{0.44} \\ \midrule
\multicolumn{1}{l}{\textbf{Leg-UP}} & 0.02 & 0.02 & 0.04 & 0.06 & 0.04 & 0.06 & 0.06 & 0.10 \\
\multicolumn{1}{l}{\textbf{\quad +Target}} & 0.10 & 0.14 & 0.14 & 0.26 & 0.12 & 0.30 & 0.14 & 0.32 \\
\multicolumn{1}{l}{\textbf{\quad + UBA(w/o $\mathcal{S}_\phi$)}} & {\ul 0.16} & {\ul 0.22} & {\ul 0.18} & {\ul 0.32} &{ \ul 0.16} & {\ul 0.36} & {\ul 0.20}& {\ul 0.36} \\
\multicolumn{1}{l}{\textbf{\quad + UBA(w/ $\mathcal{S}_\phi$)}}& \textbf{0.20} & \textbf{0.28} & \textbf{0.26} & \textbf{0.34} & \textbf{0.30} & \textbf{0.36} & \textbf{0.32} & \textbf{0.44} \\ \bottomrule
\end{tabular}
}
}
\end{center}
\vspace{-0.2cm}
\end{table}

\vspace{3pt}
\noindent\textbf{$\bullet$ Effect of the Proportions of Accessible Interactions (RQ2).} 
The proportion of accessible interactions refers to the ratio of user interactions that the attacker can access. Both the estimation of treatment effects and the generation of fake users by the attackers are using these real user interactions. 
In real-world attack scenarios, attackers can only collect a limited portion of real users' interactions. 
Therefore, using the interactions of partial users can demonstrate the practicality of the attackers. 
We demonstrate the effects of different proportions of interactions in Table~\ref{tab:data_ratio}. 
From this table, we can observe the followings: 1) as the proportion increases, the attack performance shows a growing trend, indicating that receiving more user interactions is beneficial for attackers. And 2) two UBA methods usually achieve better performance than the baselines, validating the robustness of UBA under varying proportions of accessible interactions.

\vspace{3pt}
\noindent\textbf{$\bullet$ Varying Numbers of Target Users (RQ2).} 
In Table~\ref{tab:target_user}, we evaluate the attack performance of increasing the number of target users while maintaining the same fake user budgets.  
From this table, we can observe that: 1) with increasing target user numbers, two UBA methods accomplish better attack results than the baselines, demonstrating that UBA can handle the attacks on different numbers of target users. 
2) Furthermore, two UBA methods remain relatively high hit ratios with limited fake user budgets when the number of target users is large, thereby guaranteeing attack effectiveness even if attackers seek to enlarge the attack scope. 

\begin{table}[t]
\setlength{\abovecaptionskip}{0cm}
\setlength{\belowcaptionskip}{0cm}
\caption{Performance with different numbers of target users.}
\label{tab:target_user}
\begin{center}
\setlength{\tabcolsep}{0.6mm}{
\resizebox{0.48\textwidth}{!}{
\begin{tabular}{lcccccc}
\hline
 & \multicolumn{2}{c}{\textbf{Target user 50}} & \multicolumn{2}{c}{\textbf{Target user 100}} & \multicolumn{2}{c}{\textbf{Target user 500}} \\
 & \textbf{HR@10} & \textbf{HR@20} & \textbf{HR@10} & \textbf{HR@20} & \textbf{HR@10} & \textbf{HR@20} \\ \hline
\textbf{Beofore Attack} & 0.00 & 0.02 & 0.04 & 0.06 & 0.00 & 0.02 \\
\textbf{Random Attack} & 0.00 & 0.00 & 0.10 & 0.14 & 0.10 & 0.15 \\
\textbf{Segment Attack} & 0.02 & 0.02 & 0.02 & 0.12 & 0.15 & 0.17 \\
\textbf{Bandwagon Attack} & 0.02 & 0.02 & 0.14 & 0.20 & 0.13 & 0.18 \\
\textbf{Average Attack} & 0.00 & 0.00 & 0.10 & 0.13 & 0.10 & 0.11 \\
\textbf{WGAN} & 0.02 & 0.02 & 0.05 & 0.12 & 0.12 & 0.18 \\
\textbf{DADA-DICT} & 0.04 & 0.04 & 0.31 & 0.42 & 0.30 & 0.53 \\
\textbf{DADA-DIV} & 0.00 & 0.04 & 0.20 & 0.33 & 0.29 & 0.41 \\
\textbf{DADA} & 0.06 & 0.10 & {0.30} & 0.44 & 0.28 & 0.55 \\ \hline
\textbf{AIA} & 0.02 & 0.02 & 0.18 & 0.23 & 0.17 & 0.30 \\
\textbf{+Target} & 0.20 & 0.38 & 0.20 & 0.38 & 0.25 & 0.39 \\
\textbf{+   UBA(w/o $\mathcal{S}_{\phi}$)} & {\ul 0.24} & {\ul 0.40} & {\ul 0.33} & {\ul 0.51} & {\ul 0.31} & {\ul 0.47} \\
\textbf{+   UBA(w/ $\mathcal{S}_{\phi}$)} & \textbf{0.26} & \textbf{0.44} & \textbf{0.44} & \textbf{0.70} & \textbf{0.38} & \textbf{0.55} \\ \hline
\textbf{AUSH} & 0.02 & 0.02 & 0.23 & 0.32 & 0.20 & 0.31 \\
\textbf{+Target} & 0.24 & 0.36 & 0.31 & 0.45 & 0.28 & 0.36 \\
\textbf{+   UBA(w/o $\mathcal{S}_{\phi}$)} & {\ul 0.28} & {\ul 0.40} & {\ul 0.44} & {\ul 0.56} & {\ul 0.34} & {\ul 0.39} \\
\textbf{+   UBA(w/ $\mathcal{S}_{\phi}$)} & \textbf{0.32} & \textbf{0.42} & \textbf{0.59} & \textbf{0.73} & \textbf{0.39} & \textbf{0.55} \\ \hline
\textbf{Legup} & 0.04 & 0.06 & 0.22 & 0.36 & 0.21 & 0.31 \\
\textbf{+Target} & 0.14 & 0.26 & 0.30 & 0.33 & 0.26 & 0.36 \\
\textbf{+   UBA(w/o $\mathcal{S}_{\phi}$)} & {\ul 0.18} & {\ul 0.32} & {\ul 0.41} & {\ul 0.50} & {\ul 0.40} & \textbf{0.60} \\
\textbf{+   UBA(w/ $\mathcal{S}_{\phi}$)} & \textbf{0.26} & \textbf{0.34} & \textbf{0.55} & \textbf{0.62} & \textbf{0.43} & {\ul 0.59} \\ \hline
\end{tabular}
}}
\end{center}
\vspace{-0.2cm}
\end{table}

\begin{figure}[t]
\setlength{\abovecaptionskip}{0.05cm}
\setlength{\belowcaptionskip}{-0.30cm}
\centering
\includegraphics[scale=0.3]{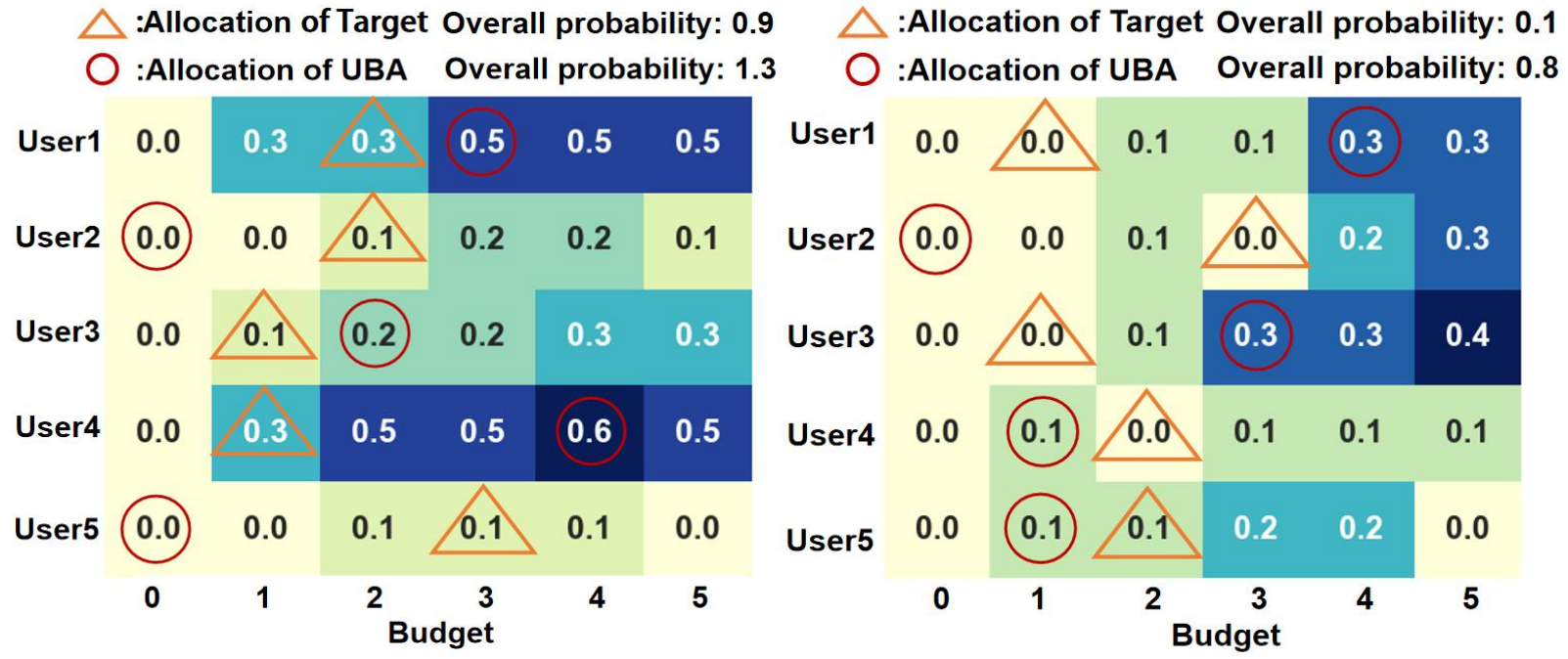}
\caption{Case study about the budget allocation on five target users. UBA allocates fake users more wisely to maximize the overall recommendation probability than Target.}
\label{fig:case_study}
\end{figure}

\vspace{3pt}
\noindent\textbf{$\bullet$ Case Study.}
To intuitively understand how UBA effectively utilizes fake user budgets to enhance attack effectiveness, we conduct a case study. We select five target users from ML-1M to compare the difference in fake user allocation between ``Target'' and UBA. Specifically, in Figure~\ref{fig:case_study}, given a target item, we show the varying recommendation probabilities of five target users under different fake user budgets, estimated by UBA(w/ $\mathcal{S}_{\phi}$). 
It is worth noting that ``Target'' randomly allocates fake user budgets while UBA maximizes overall recommendation probabilities for budget allocation. 

From Figure~\ref{fig:case_study}, we observe the following: 1) UBA allocates limited budgets to users with larger uplifts, leading to higher overall recommendation probabilities. For instance, for ``User 1'' in the left plot, increasing the budget from 2 to 3 results in an increase of 0.2 in the recommendation probability, significantly enhancing the probability of a successful attack. And 2) ``Target'', due to its random allocation of fake user budgets, usually assigns 1 to 3 fake user budgets to most users, leading to a wastage of fake user budgets on the users with constantly lower recommendation probabilities (\eg ``User 2'' and ``User 4'' in the right plot of Figure~\ref{fig:case_study}). 

\subsection{Defense Against Target User Attacks (RQ3).} 
 
In this section, we explore the ability of existing defense models against target user attackers. We examine two representative unsupervised defense models, PCA~\cite{mehta2009unsupervised} and FAP~\cite{zhang2015catch}. PCA is the most classic model to detect a group of fake users for injective attacks and FAP is a unified framework for fake user detection based on a fraudulent action propagation algorithm. These defense models usually detect fake users and exclude the detected users for recommender training. Under the defense models, we present the attack results of three backend attackers with ``Target'' and UBA on ML-1M in Table~\ref{tab:detector}.

From the table, we can observe that: 1) both PCA and FAP decrease the performance of all attackers, indicating their usefulness in defending target user attackers. In particular, FAP achieves superior defense than PCA, attributable to its proficient propagation algorithm on the user-item interaction graph. 
However, 2) even with the defense models, UBA w/ and w/o $\mathcal{S}_\phi$ show generally higher hit ratios than the vanilla attackers and ``Target''. This possibly validates the capacity of UBA to allocate budgets to target users who are more interested in the target item, making the detection of fake users challenging. 
Future research could benefit from these two observations, especially inspirations from FAP, to devise defense models specially tailored for target user attackers. 

\vspace{5pt}
\noindent$\bullet$ \textbf{More Experimental Results.}
Due to space limitations, we provide more experimental analysis in the Appendix. 
First, in addition to target user attacks, we present the attack results on all users in Section~\ref{appendix:overall_all_users}. We study the results of different victim models and defense models on all users in Section~\ref{appendix:victim_all_users} and Section~\ref{appendix:detector_all_users}. 
Besides, we only report the performance of promoting popular items in Figure~\ref{fig:budget} to save space. The results on unpopular items are in Section~\ref{appendix:unpopular_budgets}.


\begin{table}[t]
\renewcommand\arraystretch{0.9} 
\setlength{\abovecaptionskip}{0cm}
\setlength{\belowcaptionskip}{0cm}
\caption{Attack performance under two defense models.}
\label{tab:detector}
\begin{center}
\setlength{\tabcolsep}{0.6mm}{
\resizebox{0.48\textwidth}{!}{
\begin{tabular}{cccccccccc}
\toprule
\multicolumn{1}{l}{} & \multicolumn{1}{l}{} & \multicolumn{2}{c}{\textbf{AIA}} & \multicolumn{2}{c}{\textbf{+ Target}} & \multicolumn{2}{c}{\textbf{+ UBA(w/o $\mathcal{S}_\phi$)}} & \multicolumn{2}{c}{\textbf{+ UBA(w/ $\mathcal{S}_\phi$)}} \\
 &  & \textbf{Origin} & \textbf{Detector} & \textbf{Origin} & \textbf{Detector} & \textbf{Origin} & \textbf{Detector} & \textbf{Origin} & \textbf{Detector} \\ \midrule
 & \multicolumn{1}{c}{\textbf{HR@10}} & 0.02 & 0.02 & 0.20 & 0.14 & 0.26 & 0.20 & 0.24 & 0.22 \\
\multirow{-2}{*}{\textbf{PCA}} & \multicolumn{1}{c}{\textbf{HR@20}} & 0.02 & 0.02 & 0.38 & 0.28 & 0.44 & 0.32 & 0.40 & 0.32 \\ 
 & \multicolumn{1}{c}{\textbf{HR@10}} & 0.02 & 0.02 & 0.20 & 0.08 & 0.26 & 0.10 & 0.24 & 0.12 \\
\multirow{-2}{*}{\textbf{FAP}} & \multicolumn{1}{c}{\textbf{HR@20}} & 0.02 & 0.02 & 0.38 & 0.18 & 0.44 & 0.12 & 0.40 & 0.20 \\ \midrule
\multicolumn{1}{l}{} & \multicolumn{1}{l}{} & \multicolumn{2}{c}{\textbf{AUSH}} & \multicolumn{2}{c}{\textbf{+ Target}} & \multicolumn{2}{c}{\textbf{+ UBA(w/o $\mathcal{S}_\phi$)}} & \multicolumn{2}{c}{\textbf{+ UBA(w/ $\mathcal{S}_\phi$)}} \\ \midrule
 & \multicolumn{1}{c}{\textbf{HR@10}} & 0.02 & 0.02 & 0.24 & 0.20 & 0.30 & 0.18 & 0.28 & 0.22 \\
\multirow{-2}{*}{\textbf{PCA}} & \multicolumn{1}{c}{\textbf{HR@20}} & 0.02 & 0.02 & 0.36 & 0.36 & 0.42 & 0.32 & 0.40 & 0.34 \\ 
 & \multicolumn{1}{c}{\textbf{HR@10}} & 0.02 & 0.02 & 0.24 & 0.02 & 0.30 & 0.04 & 0.28 & 0.08 \\
\multirow{-2}{*}{\textbf{FAP}} & \multicolumn{1}{c}{\textbf{HR@20}} & 0.02 & 0.02 & 0.36 & 0.10 & 0.42 & 0.10 & 0.40 & 0.14 \\ \midrule
 & \multicolumn{1}{c}{} & \multicolumn{2}{c}{\textbf{Leg-UP}} & \multicolumn{2}{c}{\textbf{+ Target}} & \multicolumn{2}{c}{\textbf{+ UBA(w/o $\mathcal{S}_\phi$)}} & \multicolumn{2}{c}{\textbf{+ UBA(w/ $\mathcal{S}_\phi$)}} \\ \midrule
 & \multicolumn{1}{c}{\textbf{HR@10}} & 0.04 & 0.00 & 0.14 & 0.12 & 0.26 & 0.12 & 0.18 & 0.14 \\
\multirow{-2}{*}{\textbf{PCA}} & \multicolumn{1}{c}{\textbf{HR@20}} & 0.06 & 0.00 & 0.26 & 0.24 & 0.34 & 0.26 & 0.32 & 0.24 \\ 
 & \multicolumn{1}{c}{\textbf{HR@10}} & 0.04 & 0.00 & 0.14 & 0.04 & 0.26 & 0.06 & 0.18 & 0.08 \\
\multirow{-2}{*}{\textbf{FAP}} & \multicolumn{1}{c}{\textbf{HR@20}} & 0.06 & 0.00 & 0.26 & 0.08 & 0.34 & 0.08 & 0.32 & 0.14 \\ \bottomrule
\end{tabular}
}
}
\end{center}
\vspace{-0.4cm}
\end{table}

\section{Conclusion and Future Work}
\label{sec:conclusion}
In this work, we highlighted the significance of target user attacks and formulated the issue of varying attack difficulty across users via causal language. 
To consider the varying attack difficulty and maximize attack performance, we proposed a model-agnostic UBA framework to calculate the optimal allocation of fake user budgets. We conducted extensive experiments on three real-world datasets with diverse settings, revealing the effectiveness of UBA in performing target user attacks. 
Moreover, we validated the robustness of UBA against defense models. 
This work also emphasizes the imperative for further exploration of target user attacks and the corresponding defense strategies. 

As an initial study on uplift-enhanced recommender attack, this work leaves many promising future directions. 
First, while we have confirmed the effectiveness of two defense models to some extent, it is highly valuable to design more advanced defense strategies tailored for target user attacks. 
Second, we assume that the victim recommender models leverage CF information. Despite the popularity of CF models, UBA can be extended to attack non-CF models such as popularity-based models. 
Lastly, since large language models are gradually used for recommendation~\cite{bao2023bi,bao2023tallrec,lin2023multi}, how to attack and defend them from target user attacks deserves attention.

{
\balance
\bibliographystyle{plainnat}
\bibliography{reference}

\begin{thebibliography}{80}
\providecommand{\natexlab}[1]{#1}
\providecommand{\url}[1]{\texttt{#1}}
\expandafter\ifx\csname urlstyle\endcsname\relax
  \providecommand{\doi}[1]{doi: #1}\else
  \providecommand{\doi}{doi: \begingroup \urlstyle{rm}\Url}\fi

\bibitem[Ai et~al.(2022)Ai, Li, Gong, Yu, Xue, Zhang, Zhang, and Jiang]{Ai-LBCF-2022}
Meng Ai, Biao Li, Heyang Gong, Qingwei Yu, Shengjie Xue, Yuan Zhang, Yunzhou Zhang, and Peng Jiang.
\newblock Lbcf: A large-scale budget-constrained causal forest algorithm.
\newblock In \emph{WWW}, pages 2310--2319. ACM, 2022.

\bibitem[Aktukmak et~al.(2019)Aktukmak, Yilmaz, and Uysal]{aktukmak2019quick}
Mehmet Aktukmak, Yasin Yilmaz, and Ismail Uysal.
\newblock Quick and accurate attack detection in recommender systems through user attributes.
\newblock In \emph{RecSys}, pages 348--352. ACM, 2019.

\bibitem[Arjovsky et~al.(2017)Arjovsky, Chintala, and Bottou]{https://doi.org/10.48550/arxiv.1701.07875}
Martin Arjovsky, Soumith Chintala, and L{\'e}on Bottou.
\newblock {W}asserstein generative adversarial networks.
\newblock In Doina Precup and Yee~Whye Teh, editors, \emph{ICML}, pages 214--223. PMLR, 2017.

\bibitem[Asghar(2016)]{asghar2016yelp}
Nabiha Asghar.
\newblock Yelp dataset challenge: Review rating prediction.
\newblock \emph{arXiv preprint arXiv:1605.05362}, 2016.

\bibitem[Bao et~al.(2023{\natexlab{a}})Bao, Zhang, Wang, Zhang, Yang, Luo, Feng, He, and Tian]{bao2023bi}
Keqin Bao, Jizhi Zhang, Wenjie Wang, Yang Zhang, Zhengyi Yang, Yancheng Luo, Fuli Feng, Xiangnaan He, and Qi~Tian.
\newblock A bi-step grounding paradigm for large language models in recommendation systems.
\newblock \emph{arXiv:2308.08434}, 2023{\natexlab{a}}.

\bibitem[Bao et~al.(2023{\natexlab{b}})Bao, Zhang, Zhang, Wang, Feng, and He]{bao2023tallrec}
Keqin Bao, Jizhi Zhang, Yang Zhang, Wenjie Wang, Fuli Feng, and Xiangnan He.
\newblock Tallrec: An effective and efficient tuning framework to align large language model with recommendation.
\newblock In \emph{RecSys}. ACM, 2023{\natexlab{b}}.

\bibitem[Bhaumik et~al.(2006)Bhaumik, Williams, Mobasher, and Burke]{bhaumik2006securing}
Runa Bhaumik, Chad Williams, Bamshad Mobasher, and Robin Burke.
\newblock Securing collaborative filtering against malicious attacks through anomaly detection.
\newblock In \emph{ITWP}, page~10. AAAI, 2006.

\bibitem[Burke et~al.(2005{\natexlab{a}})Burke, Mobasher, Bhaumik, and Williams]{1565730}
R.~Burke, B.~Mobasher, R.~Bhaumik, and C.~Williams.
\newblock Segment-based injection attacks against collaborative filtering recommender systems.
\newblock In \emph{ICDM}, page~4. IEEE, 2005{\natexlab{a}}.

\bibitem[Burke et~al.(2005{\natexlab{b}})Burke, Mobasher, and Bhaumik]{burke2005limited}
Robin Burke, Bamshad Mobasher, and Runa Bhaumik.
\newblock Limited knowledge shilling attacks in collaborative filtering systems.
\newblock In \emph{IJCAI}, pages 17--24. ACM, 2005{\natexlab{b}}.

\bibitem[Cao et~al.(2013)Cao, Wu, Mao, and Zhang]{cao2013shilling}
Jie Cao, Zhiang Wu, Bo~Mao, and Yanchun Zhang.
\newblock Shilling attack detection utilizing semi-supervised learning method for collaborative recommender system.
\newblock \emph{WWW}, 16:\penalty0 729--748, 2013.

\bibitem[Cao et~al.(2020)Cao, Chen, Yao, Wang, and Zhang]{cao2020adversarial}
Yuanjiang Cao, Xiaocong Chen, Lina Yao, Xianzhi Wang, and Wei~Emma Zhang.
\newblock Adversarial attacks and detection on reinforcement learning-based interactive recommender systems.
\newblock In \emph{SIGIR}, pages 1669--1672. ACM, 2020.

\bibitem[Chen et~al.(2022)Chen, Fan, Zhu, Zhao, Yuan, Li, and Huang]{chen22ke}
Jingfan Chen, Wenqi Fan, Guanghui Zhu, Xiangyu Zhao, Chunfeng Yuan, Qing Li, and Yihua Huang.
\newblock Knowledge-enhanced black-box attacks for recommendations.
\newblock In \emph{KDD}, page 108–117. ACM, 2022.

\bibitem[Davoudi and Chatterjee(2017)]{davoudi2017detection}
Anahita Davoudi and Mainak Chatterjee.
\newblock Detection of profile injection attacks in social recommender systems using outlier analysis.
\newblock In \emph{ICBD}, pages 2714--2719. IEEE, 2017.

\bibitem[Deldjoo et~al.(2019)Deldjoo, Di~Noia, and Merra]{deldjoo2019assessing}
Yashar Deldjoo, Tommaso Di~Noia, and Felice~Antonio Merra.
\newblock Assessing the impact of a user-item collaborative attack on class of users.
\newblock \emph{arXiv:1908.07968}, 2019.

\bibitem[Deldjoo et~al.(2021)Deldjoo, Noia, and Merra]{deldjoo2021survey}
Yashar Deldjoo, Tommaso~Di Noia, and Felice~Antonio Merra.
\newblock A survey on adversarial recommender systems: from attack/defense strategies to generative adversarial networks.
\newblock \emph{CSUR}, 54\penalty0 (2):\penalty0 1--38, 2021.

\bibitem[Dou et~al.(2018)Dou, Yu, Xiong, Gao, Song, and Fang]{dou2018collaborative}
Tong Dou, Junliang Yu, Qingyu Xiong, Min Gao, Yuqi Song, and Qianqi Fang.
\newblock Collaborative shilling detection bridging factorization and user embedding.
\newblock In \emph{EAI CollaborateCom}, pages 459--469. Springer, 2018.

\bibitem[Fan et~al.(2021)Fan, Derr, Zhao, Ma, Liu, Wang, Tang, and Li]{fan2021attacking}
Wenqi Fan, Tyler Derr, Xiangyu Zhao, Yao Ma, Hui Liu, Jianping Wang, Jiliang Tang, and Qing Li.
\newblock Attacking black-box recommendations via copying cross-domain user profiles.
\newblock In \emph{ICDE}, pages 1583--1594. IEEE, 2021.

\bibitem[Fang et~al.(2018)Fang, Yang, Gong, and Liu]{fang2018poisoning}
Minghong Fang, Guolei Yang, Neil~Zhenqiang Gong, and Jia Liu.
\newblock Poisoning attacks to graph-based recommender systems.
\newblock In \emph{ACSAC}, pages 381--392. ACM, 2018.

\bibitem[Fang et~al.(2020)Fang, Gong, and Liu]{fang2020influence}
Minghong Fang, Neil~Zhenqiang Gong, and Jia Liu.
\newblock Influence function based data poisoning attacks to top-n recommender systems.
\newblock In \emph{WWW}, pages 3019--3025. ACM, 2020.

\bibitem[Gao et~al.(2020)Gao, Qi, Huang, and Sha]{gao2020shilling}
Jianling Gao, Lingtao Qi, Haiping Huang, and Chao Sha.
\newblock Shilling attack detection scheme in collaborative filtering recommendation system based on recurrent neural network.
\newblock In \emph{FICC}, pages 634--644. Springer, 2020.

\bibitem[Gutierrez and G{\'e}rardy(2017)]{Causal-gutierrez-2017}
Pierre Gutierrez and Jean-Yves G{\'e}rardy.
\newblock Causal inference and uplift modelling: A review of the literature.
\newblock In \emph{PMLR}, pages 1--13, 2017.

\bibitem[Harper and Konstan(2015)]{10.1145/2827872}
F.~Maxwell Harper and Joseph~A. Konstan.
\newblock The movielens datasets: History and context.
\newblock \emph{ACM}, 5\penalty0 (4), 2015.

\bibitem[He et~al.(2017)He, Liao, Zhang, Nie, Hu, and Chua]{He2017Neural}
Xiangnan He, Lizi Liao, Hanwang Zhang, Liqiang Nie, Xia Hu, and Tat-Seng Chua.
\newblock Neural collaborative filtering.
\newblock In \emph{WWW}, pages 173--182. ACM, 2017.

\bibitem[He et~al.(2018)He, He, Du, and Chua]{he2018adversarial}
Xiangnan He, Zhankui He, Xiaoyu Du, and Tat-Seng Chua.
\newblock Adversarial personalized ranking for recommendation.
\newblock In \emph{SIGIR}, pages 355--364. ACM, 2018.

\bibitem[He et~al.(2020)He, Deng, Wang, Li, Zhang, and Wang]{he2020lightgcn}
Xiangnan He, Kuan Deng, Xiang Wang, Yan Li, Yongdong Zhang, and Meng Wang.
\newblock Lightgcn: Simplifying and powering graph convolution network for recommendation.
\newblock In \emph{SIGIR}, pages 639--648. ACM, 2020.

\bibitem[Huang et~al.(2021)Huang, Mu, Gong, Li, Liu, and Xu]{huang2021data}
Hai Huang, Jiaming Mu, Neil~Zhenqiang Gong, Qi~Li, Bin Liu, and Mingwei Xu.
\newblock Data poisoning attacks to deep learning based recommender systems.
\newblock \emph{arXiv:2101.02644}, 2021.

\bibitem[Kaur and Goel(2016)]{kaur2016shilling}
Parneet Kaur and Shivani Goel.
\newblock Shilling attack models in recommender system.
\newblock In \emph{ICICT}, pages 1--5. IEEE, 2016.

\bibitem[Koren et~al.(2009)Koren, Bell, and Volinsky]{koren2009matrix}
Yehuda Koren, Robert Bell, and Chris Volinsky.
\newblock Matrix factorization techniques for recommender systems.
\newblock \emph{Computer}, 42\penalty0 (8):\penalty0 30--37, 2009.

\bibitem[Lam and Riedl(2004)]{lam2004shilling}
Shyong~K Lam and John Riedl.
\newblock Shilling recommender systems for fun and profit.
\newblock In \emph{WWW}, pages 393--402. ACM, 2004.

\bibitem[Li et~al.(2016{\natexlab{a}})Li, Wang, Singh, and Vorobeychik]{li2016data}
Bo~Li, Yining Wang, Aarti Singh, and Yevgeniy Vorobeychik.
\newblock Data poisoning attacks on factorization-based collaborative filtering.
\newblock \emph{NeurIPS}, 2016{\natexlab{a}}.

\bibitem[Li et~al.(2022{\natexlab{a}})Li, Di, and Chen]{li2022revisiting}
Haoyang Li, Shimin Di, and Lei Chen.
\newblock Revisiting injective attacks on recommender systems.
\newblock In \emph{NeurIPS}, 2022{\natexlab{a}}.

\bibitem[Li et~al.(2022{\natexlab{b}})Li, Di, Li, Chen, and Cao]{li2022black}
Haoyang Li, Shimin Di, Zijian Li, Lei Chen, and Jiannong Cao.
\newblock Black-box adversarial attack and defense on graph neural networks.
\newblock In \emph{ICDE}, pages 1017--1030. IEEE, 2022{\natexlab{b}}.

\bibitem[Li et~al.(2016{\natexlab{b}})Li, Gao, Li, Zeng, Xiong, and Hirokawa]{li2016shilling}
Wentao Li, Min Gao, Hua Li, Jun Zeng, Qingyu Xiong, and Sachio Hirokawa.
\newblock Shilling attack detection in recommender systems via selecting patterns analysis.
\newblock \emph{IEICE}, 99\penalty0 (10):\penalty0 2600--2611, 2016{\natexlab{b}}.

\bibitem[Lin et~al.(2020)Lin, Chen, Li, Xiao, Li, and Yang]{lin2020attacking}
Chen Lin, Si~Chen, Hui Li, Yanghua Xiao, Lianyun Li, and Qian Yang.
\newblock Attacking recommender systems with augmented user profiles.
\newblock In \emph{CIKM}, pages 855--864. ACM, 2020.

\bibitem[Lin et~al.(2022)Lin, Chen, Zeng, Zhang, Gao, and Li]{Lin_2022}
Chen Lin, Si~Chen, Meifang Zeng, Sheng Zhang, Min Gao, and Hui Li.
\newblock Shilling black-box recommender systems by learning to generate fake user profiles.
\newblock \emph{TNNLS}, pages 1--15, 2022.

\bibitem[Lin et~al.(2023)Lin, Wang, Li, Feng, Ng, and Chua]{lin2023multi}
Xinyu Lin, Wenjie Wang, Yongqi Li, Fuli Feng, See-Kiong Ng, and Tat-Seng Chua.
\newblock A multi-facet paradigm to bridge large language model and recommendation.
\newblock \emph{arXiv:2310.06491}, 2023.

\bibitem[Linden et~al.(2003)Linden, Smith, and York]{1167344}
G.~Linden, B.~Smith, and J.~York.
\newblock Amazon.com recommendations: item-to-item collaborative filtering.
\newblock \emph{IC}, 7\penalty0 (1):\penalty0 76--80, 2003.

\bibitem[Liu et~al.(2020)Liu, Xia, Chen, He, Yang, and Zheng]{liu2020certifiable}
Yang Liu, Xianzhuo Xia, Liang Chen, Xiangnan He, Carl Yang, and Zibin Zheng.
\newblock Certifiable robustness to discrete adversarial perturbations for factorization machines.
\newblock In \emph{SIGIR}, pages 419--428. ACM, 2020.

\bibitem[Liu et~al.(2022)Liu, Luo, Wu, Liu, and Li]{liu2022towards}
Zihan Liu, Yun Luo, Lirong Wu, Zicheng Liu, and Stan~Z. Li.
\newblock Towards reasonable budget allocation in untargeted graph structure attacks via gradient debias.
\newblock In \emph{NeurIPS}, 2022.

\bibitem[Mehta and Nejdl(2009)]{mehta2009unsupervised}
Bhaskar Mehta and Wolfgang Nejdl.
\newblock Unsupervised strategies for shilling detection and robust collaborative filtering.
\newblock \emph{UMUAI}, 19:\penalty0 65--97, 2009.

\bibitem[Mobasher et~al.(2007)Mobasher, Burke, Bhaumik, and Williams]{mobasher2007toward}
Bamshad Mobasher, Robin Burke, Runa Bhaumik, and Chad Williams.
\newblock Toward trustworthy recommender systems: An analysis of attack models and algorithm robustness.
\newblock \emph{TOIT}, 7\penalty0 (4):\penalty0 23--es, 2007.

\bibitem[Nguyen~Thanh et~al.(2023)Nguyen~Thanh, Quach, Nguyen, Huynh, Vu, Nguyen, Jo, and Nguyen]{Nguyen23PGRS}
Toan Nguyen~Thanh, Nguyen Duc~Khang Quach, Thanh~Tam Nguyen, Thanh~Trung Huynh, Viet~Hung Vu, Phi~Le Nguyen, Jun Jo, and Quoc Viet~Hung Nguyen.
\newblock Poisoning gnn-based recommender systems with generative surrogate-based attacks.
\newblock \emph{ACM}, 41\penalty0 (3), feb 2023.

\bibitem[O'Mahony et~al.(2005)O'Mahony, Hurley, and Silvestre]{o2005recommender}
Michael~P O'Mahony, Neil~J Hurley, and Gu{\'e}nol{\'e}~CM Silvestre.
\newblock Recommender systems: Attack types and strategies.
\newblock In \emph{AAAI}, pages 334--339. AAAI, 2005.

\bibitem[Pang et~al.(2018)Pang, Gao, Tao, and Zhou]{pang2018unorganized}
Ming Pang, Wei Gao, Min Tao, and Zhi-Hua Zhou.
\newblock Unorganized malicious attacks detection.
\newblock \emph{NeurIPS}, 2018.

\bibitem[Paudel et~al.(2016)Paudel, Christoffel, Newell, and Bernstein]{Paudel16Up}
Bibek Paudel, Fabian Christoffel, Chris Newell, and Abraham Bernstein.
\newblock Updatable, accurate, diverse, and scalable recommendations for interactive applications.
\newblock \emph{ACM}, 7\penalty0 (1), dec 2016.

\bibitem[Rappaz et~al.(2017)Rappaz, Vladarean, McAuley, and Catasta]{10.1145/3018661.3018696}
J\'{e}r\'{e}mie Rappaz, Maria-Luiza Vladarean, Julian McAuley, and Michele Catasta.
\newblock Bartering books to beers: A recommender system for exchange platforms.
\newblock In \emph{WSDM}, pages 505--514. ACM, 2017.

\bibitem[Shahrasbi et~al.(2020)Shahrasbi, Mani, Arrabothu, Sharma, Achan, and Kumar]{DBLP:journals/corr/abs-2012-02509}
Behzad Shahrasbi, Venugopal Mani, Apoorv~Reddy Arrabothu, Deepthi Sharma, Kannan Achan, and Sushant Kumar.
\newblock On detecting data pollution attacks on recommender systems using sequential gans.
\newblock In \emph{arXiv:2012.02509}, 2020.

\bibitem[Sharma and Gera(2013)]{sharma2013survey}
Lalita Sharma and Anju Gera.
\newblock A survey of recommendation system: Research challenges.
\newblock \emph{IJETT}, 4\penalty0 (5):\penalty0 1989--1992, 2013.

\bibitem[Song et~al.(2020)Song, Li, Hu, Wu, Li, Li, and Gao]{song2020poisonrec}
Junshuai Song, Zhao Li, Zehong Hu, Yucheng Wu, Zhenpeng Li, Jian Li, and Jun Gao.
\newblock Poisonrec: an adaptive data poisoning framework for attacking black-box recommender systems.
\newblock In \emph{ICDE}, pages 157--168. IEEE, 2020.

\bibitem[Sun et~al.(2022)Sun, Wang, Jing, Cui, Song, and Nie]{sun2022counterfactual}
Teng Sun, Wenjie Wang, Liqaing Jing, Yiran Cui, Xuemeng Song, and Liqiang Nie.
\newblock Counterfactual reasoning for out-of-distribution multimodal sentiment analysis.
\newblock In \emph{MM}, pages 15--23. ACM, 2022.

\bibitem[Tang et~al.(2020)Tang, Wen, and Wang]{tang2020revisiting}
Jiaxi Tang, Hongyi Wen, and Ke~Wang.
\newblock Revisiting adversarially learned injection attacks against recommender systems.
\newblock In \emph{RecSys}, pages 318--327. ACM, 2020.

\bibitem[Tang et~al.(2019)Tang, Du, He, Yuan, Tian, and Chua]{tang2019adversarial}
Jinhui Tang, Xiaoyu Du, Xiangnan He, Fajie Yuan, Qi~Tian, and Tat-Seng Chua.
\newblock Adversarial training towards robust multimedia recommender system.
\newblock \emph{TKDE}, 32\penalty0 (5):\penalty0 855--867, 2019.

\bibitem[Tu et~al.(2021)Tu, Basu, DiCiccio, Bansal, Nandy, Jaikumar, and Chatterjee]{Tu-Personalized-2021}
Ye~Tu, Kinjal Basu, Cyrus DiCiccio, Romil Bansal, Preetam Nandy, Padmini Jaikumar, and Shaunak Chatterjee.
\newblock Personalized treatment selection using causal heterogeneity.
\newblock In \emph{WWW}, pages 1574--1585. ACM, 2021.

\bibitem[Wilson and Seminario(2013)]{wilson2013power}
David~C Wilson and Carlos~E Seminario.
\newblock When power users attack: assessing impacts in collaborative recommender systems.
\newblock In \emph{RecSys}, pages 427--430, 2013.

\bibitem[Wu et~al.(2021{\natexlab{a}})Wu, Lian, Ge, Zhu, and Chen]{wu-Triple-2021}
Chenwang Wu, Defu Lian, Yong Ge, Zhihao Zhu, and Enhong Chen.
\newblock Triple adversarial learning for influence based poisoning attack in recommender systems.
\newblock In \emph{KDD}, pages 1830--1840. ACM, 2021{\natexlab{a}}.

\bibitem[Wu et~al.(2021{\natexlab{b}})Wu, Lian, Ge, Zhu, and Chen]{wu21triple}
Chenwang Wu, Defu Lian, Yong Ge, Zhihao Zhu, and Enhong Chen.
\newblock Triple adversarial learning for influence based poisoning attack in recommender systems.
\newblock In \emph{KDD}, page 1830–1840. ACM, 2021{\natexlab{b}}.

\bibitem[Wu et~al.(2021{\natexlab{c}})Wu, Lian, Ge, Zhu, Chen, and Yuan]{wu2021fight}
Chenwang Wu, Defu Lian, Yong Ge, Zhihao Zhu, Enhong Chen, and Senchao Yuan.
\newblock Fight fire with fire: towards robust recommender systems via adversarial poisoning training.
\newblock In \emph{SIGIR}, pages 1074--1083. ACM, 2021{\natexlab{c}}.

\bibitem[Wu et~al.(2021{\natexlab{d}})Wu, Lian, Ge, Zhu, Chen, and Yuan]{wu21Fight}
Chenwang Wu, Defu Lian, Yong Ge, Zhihao Zhu, Enhong Chen, and Senchao Yuan.
\newblock Fight fire with fire: Towards robust recommender systems via adversarial poisoning training.
\newblock In \emph{SIGIR}, page 1074–1083. ACM, 2021{\natexlab{d}}.

\bibitem[Wu et~al.(2022)Wu, Li, Deng, Hu, Dai, Dong, Sun, Zhang, and Zhou]{wu2022opportunity}
Peng Wu, Haoxuan Li, Yuhao Deng, Wenjie Hu, Quanyu Dai, Zhenhua Dong, Jie Sun, Rui Zhang, and Xiao-Hua Zhou.
\newblock On the opportunity of causal learning in recommendation systems: Foundation, estimation, prediction and challenges.
\newblock In \emph{IJCAI}, pages 23--29, 2022.

\bibitem[Wu et~al.(2011)Wu, Cao, Mao, and Wang]{wu2011semi}
Zhiang Wu, Jie Cao, Bo~Mao, and Youquan Wang.
\newblock Semi-sad: applying semi-supervised learning to shilling attack detection.
\newblock In \emph{RecSys}, pages 289--292. ACM, 2011.

\bibitem[Xie et~al.(2021)Xie, Liu, Wu, Sun, Liu, Chen, Gao, Cui, and Ding]{xie21causCF}
Xu~Xie, Zhaoyang Liu, Shiwen Wu, Fei Sun, Cihang Liu, Jiawei Chen, Jinyang Gao, Bin Cui, and Bolin Ding.
\newblock Causcf: Causal collaborative filtering for recommendation effect estimation.
\newblock In \emph{CIKM}, page 4253–4263. ACM, 2021.

\bibitem[Xing et~al.(2013{\natexlab{a}})Xing, Meng, Doozan, Snoeren, Feamster, and Lee]{182952}
Xingyu Xing, Wei Meng, Dan Doozan, Alex~C. Snoeren, Nick Feamster, and Wenke Lee.
\newblock Take this personally: Pollution attacks on personalized services.
\newblock In \emph{USENIX}, pages 671--686. USENIX, 2013{\natexlab{a}}.

\bibitem[Xing et~al.(2013{\natexlab{b}})Xing, Meng, Doozan, Snoeren, Feamster, and Lee]{xing2013take}
Xinyu Xing, Wei Meng, Dan Doozan, Alex~C Snoeren, Nick Feamster, and Wenke Lee.
\newblock Take this personally: Pollution attacks on personalized services.
\newblock In \emph{USENIX}, pages 671--686, 2013{\natexlab{b}}.

\bibitem[Yang et~al.(2018)Yang, Gao, Yu, Song, and Wang]{yang2018detection}
Fan Yang, Min Gao, Junliang Yu, Yuqi Song, and Xinyi Wang.
\newblock Detection of shilling attack based on bayesian model and user embedding.
\newblock In \emph{ICTAI}, pages 639--646. IEEE, 2018.

\bibitem[Yang et~al.(2017)Yang, Gong, and Cai]{yang2017fake}
Guolei Yang, Neil~Zhenqiang Gong, and Ying Cai.
\newblock Fake co-visitation injection attacks to recommender systems.
\newblock In \emph{NDSS}, 2017.

\bibitem[Yao et~al.(2021)Yao, Chu, Li, Li, Gao, and Zhang]{yao21survey}
Liuyi Yao, Zhixuan Chu, Sheng Li, Yaliang Li, Jing Gao, and Aidong Zhang.
\newblock A survey on causal inference.
\newblock \emph{ACM}, may 2021.

\bibitem[You et~al.(2023)You, Li, Ding, Zhang, Feng, Pan, and Yang]{you23anti}
Xiaoyu You, Chi Li, Daizong Ding, Mi~Zhang, Fuli Feng, Xudong Pan, and Min Yang.
\newblock Anti-fakeu: Defending shilling attacks on graph neural network based recommender model.
\newblock In \emph{WWW}, page 938–948. ACM, 2023.

\bibitem[Yu et~al.(2021)Yu, Zheng, Xu, Ma, Gao, and Zhang]{9498095}
Hongtao Yu, Haihong Zheng, Yishu Xu, Ru~Ma, Dingli Gao, and Fuzhi Zhang.
\newblock Detecting group shilling attacks in recommender systems based on maximum dense subtensor mining.
\newblock In \emph{ICAICA}, pages 644--648. IEEE, 2021.

\bibitem[Yu et~al.(2022)Yu, Liu, Wan, Zheng, Li, and Zhou]{yu-MDP2-2022}
Sizhe Yu, Ziyi Liu, Shixiang Wan, Jia Zheng, Zang Li, and Fan Zhou.
\newblock Mdp2 forest: A constrained continuous multi-dimensional policy optimization approach for short-video recommendation.
\newblock In \emph{KDD}, pages 2388--2398. ACM, 2022.

\bibitem[Yue et~al.(2021)Yue, He, Zeng, and McAuley]{yue21bbattack}
Zhenrui Yue, Zhankui He, Huimin Zeng, and Julian McAuley.
\newblock Black-box attacks on sequential recommenders via data-free model extraction.
\newblock In \emph{RecSys}, page 44–54. ACM, 2021.

\bibitem[Zar(1972)]{spearman}
Jerrold~H Zar.
\newblock Significance testing of the \text{Spearman} rank correlation coefficient.
\newblock \emph{JASA}, 67\penalty0 (339):\penalty0 578--580, 1972.

\bibitem[Zhang et~al.(2020{\natexlab{a}})Zhang, Li, Ding, and Gao]{zhang2020practical}
Hengtong Zhang, Yaliang Li, Bolin Ding, and Jing Gao.
\newblock Practical data poisoning attack against next-item recommendation.
\newblock In \emph{WWW}, pages 2458--2464. ACM, 2020{\natexlab{a}}.

\bibitem[Zhang et~al.(2021{\natexlab{a}})Zhang, Tian, Li, Su, Yang, Zhao, and Gao]{zhang2021data}
Hengtong Zhang, Changxin Tian, Yaliang Li, Lu~Su, Nan Yang, Wayne~Xin Zhao, and Jing Gao.
\newblock Data poisoning attack against recommender system using incomplete and perturbed data.
\newblock In \emph{KDD}, pages 2154--2164. ACM, 2021{\natexlab{a}}.

\bibitem[Zhang et~al.(2021{\natexlab{b}})Zhang, Yin, Chen, Huang, Cui, and Zhang]{10.1145/3442381.3449813}
Shijie Zhang, Hongzhi Yin, Tong Chen, Zi~Huang, Lizhen Cui, and Xiangliang Zhang.
\newblock Graph embedding for recommendation against attribute inference attacks.
\newblock In \emph{www}, page 3002–3014. ACM, 2021{\natexlab{b}}.

\bibitem[Zhang et~al.(2020{\natexlab{b}})Zhang, Sheng, Alhazmi, and Li]{zhang2020adversarial}
Wei~Emma Zhang, Quan~Z Sheng, Ahoud Alhazmi, and Chenliang Li.
\newblock Adversarial attacks on deep-learning models in natural language processing: A survey.
\newblock \emph{TIST}, 11\penalty0 (3):\penalty0 1--41, 2020{\natexlab{b}}.

\bibitem[Zhang et~al.(2021{\natexlab{c}})Zhang, Li, and Liu]{zhang_unified_2021}
Weijia Zhang, Jiuyong Li, and Lin Liu.
\newblock A unified survey of treatment effect heterogeneity modelling and uplift modelling.
\newblock \emph{ACM Comput. Surv.}, 54\penalty0 (8), 2021{\natexlab{c}}.

\bibitem[Zhang et~al.(2022)Zhang, Wang, Zhao, and Wang]{zhang2022targeted}
Xudong Zhang, Zan Wang, Jingke Zhao, and Lanjun Wang.
\newblock Targeted data poisoning attack on news recommendation system.
\newblock In \emph{arXiv:2203.03560}, 2022.

\bibitem[Zhang et~al.(2015)Zhang, Tan, Zhang, Liu, Chua, and Ma]{zhang2015catch}
Yongfeng Zhang, Yunzhi Tan, Min Zhang, Yiqun Liu, Tat-Seng Chua, and Shaoping Ma.
\newblock Catch the black sheep: unified framework for shilling attack detection based on fraudulent action propagation.
\newblock In \emph{IJCAI}. ACM, 2015.

\bibitem[Zhou et~al.(2016)Zhou, Wen, Xiong, Gao, and Zeng]{zhou2016svm}
Wei Zhou, Junhao Wen, Qingyu Xiong, Min Gao, and Jun Zeng.
\newblock Svm-tia a shilling attack detection method based on svm and target item analysis in recommender systems.
\newblock \emph{Neurocomputing}, 210:\penalty0 197--205, 2016.

\bibitem[Zhu et~al.(2013)Zhu, Li, Ren, Zhou, and Xiong]{zhu2013differential}
Tianqing Zhu, Gang Li, Yongli Ren, Wanlei Zhou, and Ping Xiong.
\newblock Differential privacy for neighborhood-based collaborative filtering.
\newblock In \emph{ASONAM}, pages 752--759, 2013.

\end{thebibliography}
}

\newpage
\appendix
{
\large
\textbf{Appendix}
}
\section{Method}
We validate the correlation of Proposition~\ref{pro:A3_path} via experiments in Section~\ref{appendix:correlation_exp}, and present the theoretical analysis of Proposition~\ref{pro:A3_path} and Proposition~\ref{pro:A3_path_user} in Section~\ref{appendix:P_1_proofs} and Section~\ref{appendix:P_2_proofs}, respectively. Lastly, we detail the dynamic programming algorithm for treatment selection in Section~\ref{appendix:DP}. 

\subsection{Correlation Analysis via Experiments}\label{appendix:correlation_exp}

\begin{figure}[h]
\setlength{\abovecaptionskip}{0cm}
\setlength{\belowcaptionskip}{0cm}
\centering
\includegraphics[scale=0.2]{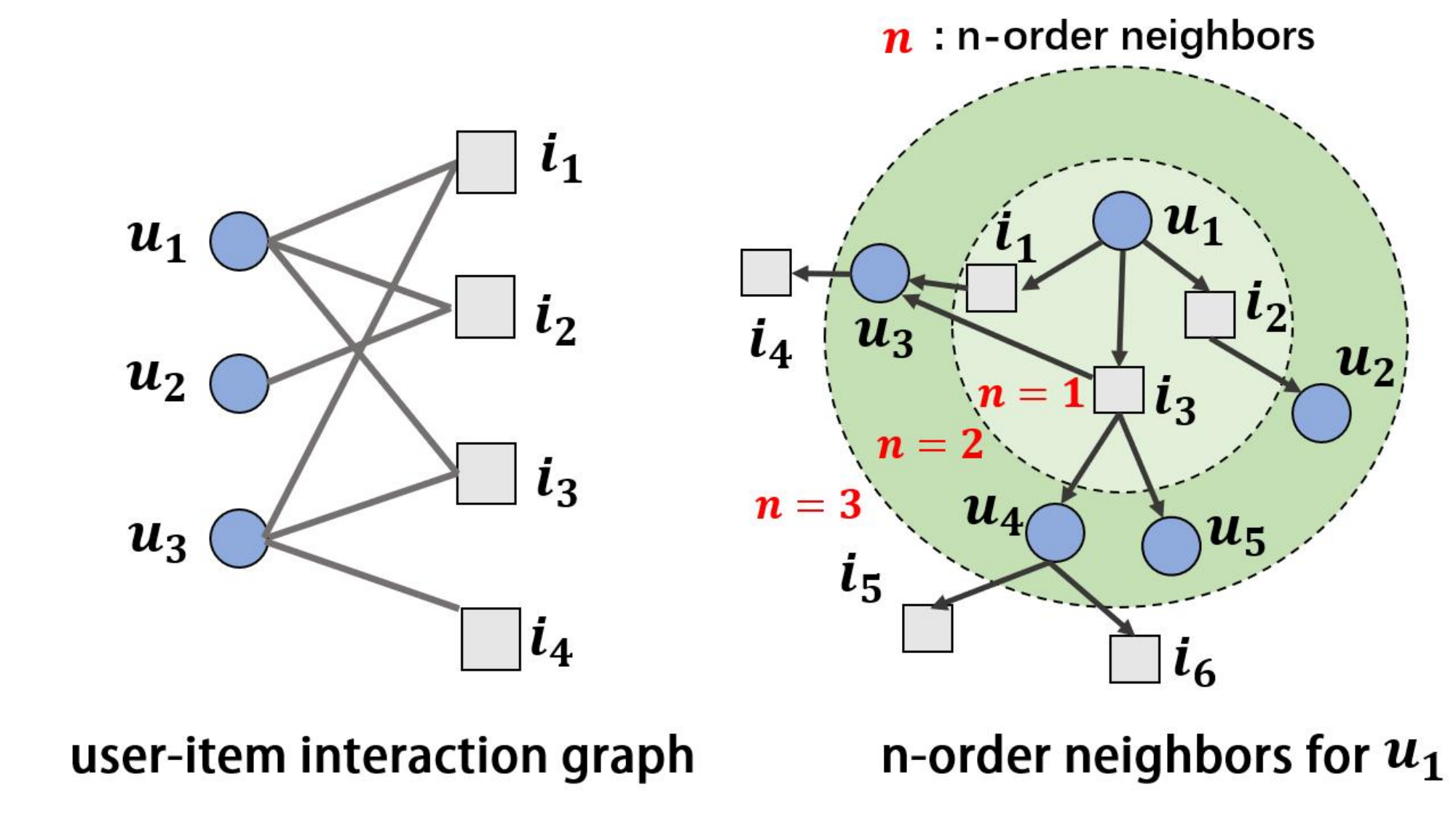}
\caption{Illustration of three-order neighbors on the user-item interaction graph.}
\label{fig:three-order}
\end{figure}

\begin{figure*}[t]
\setlength{\abovecaptionskip}{0.1cm}
\setlength{\belowcaptionskip}{0cm}
\centering
\includegraphics[scale=0.68]{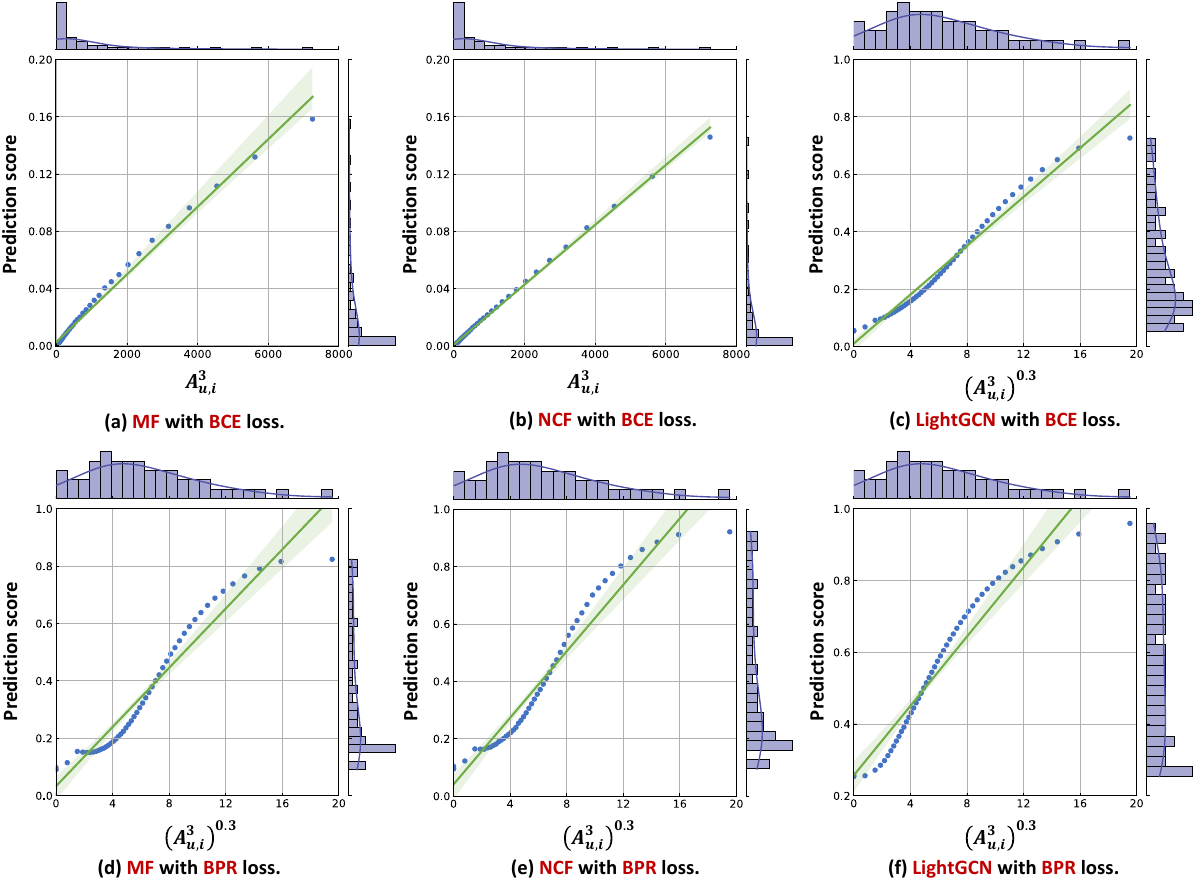}
\caption{Visualization of the correlation between the three-order path number and the normalized prediction scores from three representative CF models (\ie MF, NCF, LightGCN) with two widely used loss functions (\ie BCE and BPR). We conduct the Spearman Rank Correlation Test~\cite{spearman} for each figure. The test results with $r\geq 0.998\approx 1$ and $p\leq 1.46e^{-59}\ll0.001$ indicate the strong positive correlations in these experiments. }
\label{fig:correlation_bias_models}
\end{figure*}

\begin{figure*}[ht]
\setlength{\abovecaptionskip}{0.1cm}
\setlength{\belowcaptionskip}{0cm}
\centering
\includegraphics[scale=0.68]{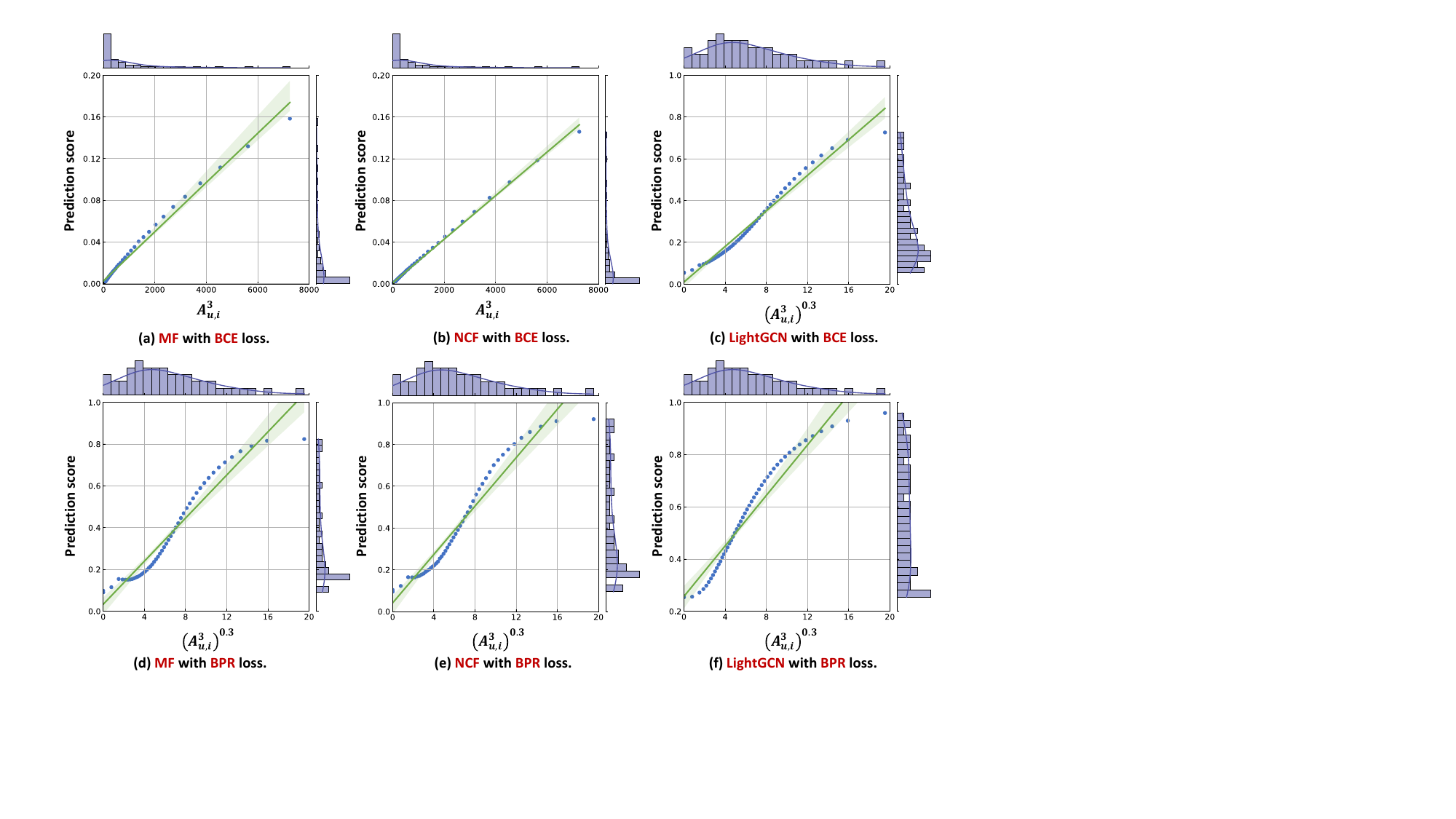}
\caption{Visualization of the correlation between the high-order path numbers and the prediction scores, where we vary the order number in $\{3,5,7\}$. The figures validate the robustness of such correlation in different order numbers. We conduct the Spearman Rank Correlation Test~\cite{spearman} for each figure. The test results with $r\geq 0.998\approx 1$ and $p\leq 1.46e^{-59}\ll0.001$ indicate the strong positive correlations in these experiments. }
\label{fig:correlation_bias_orders}
\end{figure*}

We conduct many experiments to verify the robustness of Proposition~\ref{pro:A3_path} on multiple CF models with two representative loss functions, BCE and BPR. In detail, we collect all user-item pairs without interactions in the training data and rank them via their high-order path numbers (see illustration in Figure~\ref{fig:three-order}), which are calculated via matrix multiplication over $\bm{A}$. Besides, we obtain the prediction scores of these user-item pairs by different CF models. 
The ranked user-item pairs are split into 50 groups and we report the average prediction score and high-order path number of each group by scatter plots as shown in Figure~\ref{fig:correlation_bias_models} and Figure~\ref{fig:correlation_bias_orders}.

In particular, Figure~\ref{fig:correlation_bias_models} shows the robustness of the correlation between the three-order path number and the prediction score on three CF models (\ie MF, NCF, and LightGCN) with two popular recommender loss functions (\ie BCE and BPR). Furthermore, we validate that such correlation also holds on different orders such as five-order and seven-order path numbers as depicted in Figure~\ref{fig:correlation_bias_orders}. Besides, we conduct the Spearman Rank Correlation Test~\cite{spearman} for each figure. The correlation coefficients are all larger than $0.998$ and the $p$-values are smaller than $1.46e^{-59}\ll0.001$, indicating the strong positive correlations in these experiments. 
Last but not least, in Figure~\ref{fig:correlation_bias_models}, we normalize the prediction scores via sigmoid for better visualization. Accordingly, we re-scale the three-order path numbers in some figures by $(\bm{A}^3_{u,i})^{0.3}$, where similar trends can also be obtained by the log function. 
From Figure~\ref{fig:correlation_bias_models}, we can find that the correlation between the sigmoid-normalized prediction scores and the scaled three-order path numbers is approximately linear on LightGCN and the models with BPR loss. This shows the generality and robustness of using $\alpha\cdot(\bm{A}^3_{u,i})^\beta$ to approximate the treatment effect $Y^{\theta^*}_{u,i}(\cdot)$ in Section~\ref{sec:treatment_estimation}.

\subsection{Theoretical Analysis of Proposition 1}\label{appendix:P_1_proofs}

\textbf{Reminder of Proposition 1.} Given a user and an item without historical interaction, their 
prediction score by a CF model is positively correlated with their three-order path number in $\bm{A}^3$.

\begin{proof}
    Formally, we reformulate our proposition as follows: \par 
    
    For different victim recommender models $\mathcal{M}_\theta$ with different loss functions $\mathcal{L}$, after sufficient training of the model parameters $\theta$, we have the positive correlation between the prediction score $\mathcal{M}_{\theta^*}(u,i)$ of a specific user-item pair $(u,i)$ and their three-order path number $\bm{A}^3_{u,i}$:
    \begin{equation}
        \mathcal{M}_{\theta^*}(u,i) \propto \bm{A}_{u,i}^3,
        \label{reformulated_propoA2}
    \end{equation}
    where $\theta^* = \arg\min_{\theta}\mathcal{L}(\mathcal{M}_\theta, \bm{D})$. \par
    
    To begin with, we will prove our proposition via a specific example with MF and BPR loss function, and then demonstrate how our conclusion can generalize to other CF models and loss. To prove the proposition via the specific example, we first delve into the optimization process of the model, which can be formulated as:
    \begin{equation}
    \begin{aligned}
        &\min_{\theta} \mathcal{L}_{\text{BPR}}(\mathcal{M}_\theta, \bm{D})\\ 
        &=  \frac{1}{|\bm{P}|} \sum_{(u,i)\in\bm{P}}  \frac{1}{|\mathcal{I}|} \sum_{j\in \mathcal{I}} -\log(\sigma(\mathcal{M}_\theta(u,i) - \mathcal{M}_\theta(u,j))),
    \end{aligned}
    \end{equation}
    where $\bm{P}$ denotes the positive interaction (positive user-item pair) set of $\bm{D}$. $|\bm{P}|$ represents the size of $\bm{P}$ and $\sigma(t) = 1/(1+\exp(-t))$. When we utilize MF as the recommender model, we have the prediction score $\mathcal{M}_\theta(u,i)=(\bm{e}_u^\theta)^T(\bm{e}_i^\theta)$, where the $\bm{e}_u^\theta$ and $\bm{e}_u^\theta$ denotes the learned embeddings of user $u$ and item $i$, respectively. \par
    
    Taking the partial derivative of $\mathcal{L}_{\text{BPR}}$ with respect to the embedding of user $u^*$, we get:
    \begin{equation}
    \begin{aligned}
        \frac{\partial{ \mathcal{L}_{\text{BPR}}(\mathcal{M}_\theta, \bm{D})}}{\partial{\bm{e}_{u^*}^\theta}}
        &= \frac{1}{|\bm{P}|} \sum_{i\in \mathcal{N}_{u^*}}  \frac{1}{|\mathcal{I}|} \sum_{j\in \mathcal{I}} \phi_\theta(u^*,i,j) \cdot (\bm{e}_i^\theta - \bm{e}_j^\theta) \\ 
        & \approx \frac{1}{|\bm{P}|} \sum_{i\in \mathcal{N}_{u^*}} \phi_\theta(u^*,i,j) \cdot \bm{e}_i^\theta \\
        &- \frac{1}{|\bm{P}|} \frac{|\mathcal{N}_{u^*}|}{|\mathcal{I}|}  \sum_{j\in \mathcal{I}} \phi_\theta(u^*,i,j) \cdot \bm{e}_j^\theta,
    \end{aligned}
    \label{equation:partial_derivative_u_bpr}
    \end{equation}
    where $\mathcal{N}_{u^*}$ denotes the first-order neiborhood set of user $u^*$ and $|\mathcal{N}_{u^*}|$ represents its size. We utilize function $\phi_\theta(u,i,j)$ to represent $- \left[1- \sigma(\mathcal{M}_\theta(u,i) - \mathcal{M}_\theta(u,j))\right]$, which is assumed approximately as the same value for different $(u,i,j)$ pairs. And $\phi_\theta(u,i,j)$ becomes close to zero as the training progresses. The reason for this approximation is that we assume that the MF model can well memorize or fit the training data. 
    \par

    Similarly, we obtain the partial derivative of the loss function $\mathcal{L}_{\text{BPR}}$ with respect to the embedding of item $i^*$ as follows:
    \begin{equation}
    \begin{aligned}
        \frac{\partial{ \mathcal{L}_{\text{BPR}}(\mathcal{M}_\theta, \bm{D})}}{\partial{\bm{e}_{i^*}^\theta}} &= \frac{1}{|\bm{P}|} \sum_{u\in \mathcal{N}_{i^*}} \frac{1}{|\mathcal{I}|} \sum_{j\in\mathcal{I}} \phi_\theta(u,i^*,j) \cdot \bm{e}_u^\theta \\
        &- \frac{1}{|\bm{P}|}  \frac{1}{|\mathcal{I}|} \sum_{(u',i) \in \bm{P}} \phi_\theta(u',i,i^*) \cdot\bm{e}_{u'}^\theta \\
        & \approx \frac{1}{|\bm{P}|} \sum_{u\in \mathcal{N}_{i^*}} \phi_\theta(u,i^*,j) \cdot \bm{e}_u^\theta \\
        &- \frac{1}{|\bm{P}|}   \frac{1}{|\mathcal{I}|} \sum_{u' \in \mathcal{U}} |\mathcal{N}_{u'}| \cdot \phi_\theta(u',i,i^*) \cdot\bm{e}_{u'}^\theta,
    \end{aligned}   
    \label{equation:partial_derivative_i_bpr}
    \end{equation}
     where $\mathcal{N}_{i^*}$ denotes the first-order neighborhood set of item $i^*$ and $|\mathcal{N}_{u'}|$ denotes the size of $\mathcal{N}_{u'}$. Similarly, We utilize function $\phi_\theta(u,i,j)$ to represent $- \left[1- \sigma(\mathcal{M}_\theta(u,i) - \mathcal{M}_\theta(u,j))\right]$ and assume its values on different $(u,i,j)$ pairs are  approximately becoming closer as the training processes. \par

     Based on the above derivation, we observe that the gradients of $\bm{e}_{u^*}^\theta$ and $\bm{e}_{i^*}^\theta$ always point towards a fixed target during the iteration, and their magnitudes are affected by $|\phi_\theta(u,i,j)|$, which gradually becomes smaller as the training converges. Therefore, we can approximate the converged values of  $\bm{e}_{u^*}^{\theta^*}$ and $\bm{e}_{i^*}^{\theta^*}$ as:
     \begin{equation}
         \bm{e}_{u^*}^{\theta^*} \approx  C_1\cdot \left[  \sum_{i\in \mathcal{N}_{u^*}} \bm{e}_i^{\theta^*} - \frac{|\mathcal{N}_{u^*}|}{|\mathcal{I}|}  \sum_{j\in \mathcal{I}} \bm{e}_j^{\theta^*} \right],
     \end{equation}
     \begin{equation}
         \bm{e}_{i^*}^{\theta^*} \approx C_1\cdot \left[\sum_{u\in \mathcal{N}_{i^*}} \bm{e}_u^{\theta^*} -   \frac{1}{|\mathcal{I}|} \sum_{u' \in \mathcal{U}} |\mathcal{N}_{u'}| \cdot\bm{e}_{u'}^{\theta^*} \right],
     \end{equation}
     where $C_1$ is a fixed constant. Furthermore, by using the average number of first-order neighbors $\bar{|\mathcal{N}|}$ of all nodes in the graph to approximate $|\mathcal{N}_{u'}|$ and $|\mathcal{N}_{i^*}|$, we can make the following assumption:
     \begin{assumption}
         When the model converges, the embeddings of user $u$ and item $i$ have the following form: 
         \begin{equation}
             \bm{e}_{u^*}^{\theta^*} \approx  C_1\cdot \left[  \sum_{i\in \mathcal{N}_{u^*}} \bm{e}_i^{\theta^*} - \frac{\bar{|\mathcal{N}|}}{|\mathcal{I}|}  \sum_{j\in \mathcal{I}} \bm{e}_j^{\theta^*} \right]
         \end{equation}
         \begin{equation}
             \bm{e}_{i^*}^{\theta^*} \approx C_1\cdot \left[\sum_{u\in \mathcal{N}_{i^*}} \bm{e}_u^{\theta^*} -   \frac{\bar{|\mathcal{N}|}}{|\mathcal{I}|} \sum_{u' \in \mathcal{U}} \bm{e}_{u'}^{\theta^*} \right],
         \end{equation}
         where $C_1$ is a fixed constant, $\bar{|\mathcal{N}|}$ denotes the average number of first-order neighbors of all nodes in the graph.
         \label{assumption:1}
     \end{assumption}
     Thus, we have proved the form of embeddings of users and items when the model converges. To complete the remaining proof, we need some additional assumptions.
     \begin{assumption}
         We assume that the collected user-item interactions for recommender training are sparse, which means that $\frac{\bar{|\mathcal{N}|^3}}{|\mathcal{I}|} \ll 1$.
         \label{assumption:2}
     \end{assumption}
     \begin{assumption}
         When the model converges, the user-item pairs with historical interactions have larger prediction scores, while the user-item pairs without historical interactions have smaller predicted scores. This can be formulated as: 
         \begin{align}
             (\bm{e}_{u^*}^{\theta^*})^T (\bm{e}_{i^*}^{\theta^*}) = \begin{cases}
                 C_{\text{large}}, & \text{if } (u,i)\in\bm{P} \\
                 C_{\text{small}}, & \text{if } (u,i)\notin\bm{P}
             \end{cases},
         \end{align}
         where $\bm{P}$ denotes the positive interaction (positive user-item pair) set of $\bm{D}$ and $C_{\text{small}} \ll C_{\text{large}}$.
         \label{assumption:3}
     \end{assumption}
     The Assumptions~\ref{assumption:2} and~\ref{assumption:3} above are reasonable in recommender systems since the user interactions are usually sparse and neural CF models can well fit the training interactions. Hence, we could further derive the prediction score of a user-item pair when the model converges:
     \begin{equation}
     \begin{aligned}
        &\mathcal{M}_{\theta^*}(u^*,i^*) =(\bm{e}_{u^*}^{\theta^*})^T (\bm{e}_{i^*}^{\theta^*}) \\
        & \approx  (C_1)^2 \cdot \left[  \sum_{i\in \mathcal{N}_{u^*}} \bm{e}_i^{\theta^*} - \frac{\bar{|\mathcal{N}|}}{|\mathcal{I}|}  \sum_{j\in \mathcal{I}} \bm{e}_j^{\theta^*} \right]^T \left[\sum_{u\in \mathcal{N}_{i^*}} \bm{e}_u^{\theta^*} -   \frac{\bar{|\mathcal{N}|}}{|\mathcal{I}|} \sum_{u' \in \mathcal{U}} \bm{e}_{u'}^{\theta^*} \right] \\
        & \approx (C_1)^2 \cdot [ \sum_{i\in \mathcal{N}_{u^*}}  \sum_{u\in \mathcal{N}_{i^*}} (\bm{e}_u^{\theta^*})^T (\bm{e}_i^{\theta^*}) - \frac{\bar{|\mathcal{N}|}}{|\mathcal{I}|}  \sum_{u\in \mathcal{N}_{i^*}} \sum_{j\in \mathcal{I}} (\bm{e}_j^{\theta^*})^T(\bm{e}_u^{\theta^*}) \\ 
        & - \frac{\bar{|\mathcal{N}|}}{|\mathcal{I}|} \sum_{u' \in \mathcal{U}} \sum_{i\in \mathcal{N}_{u^*}} (\bm{e}_{u'}^{\theta^*})^T(\bm{e}_i^{\theta^*}) + \frac{\bar{|\mathcal{N}|}^2}{|\mathcal{I}|^2} \sum_{u' \in \mathcal{U}} \sum_{j\in \mathcal{I}} (\bm{e}_{u'}^{\theta^*})^T(\bm{e}_j^{\theta^*}) ].
    \end{aligned}
    \label{proof_M}
     \end{equation}
     
     Notably that
     \begin{equation}
     \begin{aligned}
         & \sum_{i\in \mathcal{N}_{u^*}}  \sum_{u\in \mathcal{N}_{i^*}} (\bm{e}_u^{\theta^*})^T (\bm{e}_i^{\theta^*}) \\
         = & \bm{A}_{u,i}^3 \cdot C_{\text{large}} + (|\mathcal{N}_{u^*}||\mathcal{N}_{i^*}| - \bm{A}_{u,i}^3) \cdot C_{\text{small}} \\
         \approx & \bm{A}_{u,i}^3 \cdot C_{\text{large}}.
     \end{aligned}
     \end{equation}
     
     Hence, by neglecting infinitesimal terms, the Eq.~\eqref{proof_M} can be reformulated as:
     \begin{equation}
         \begin{aligned}
             \mathcal{M}_{\theta^*}(u^*,i^*) &\approx (C_1)^2 \cdot \left[\bm{A}_{u,i}^3 \cdot C_{\text{large}} - 2\cdot \frac{\bar{|\mathcal{N}|}^2}{|\mathcal{I}|}\cdot C_{\text{large}} +  \frac{\bar{|\mathcal{N}|}^3}{|\mathcal{I}|}\cdot C_{\text{large}} \right] \\
             & \approx (C_1)^2 \cdot \bm{A}_{u,i}^3 \cdot C_{\text{large}} \\
             & \propto \bm{A}_{u,i}^3.
         \end{aligned}
     \end{equation}
    So far, we have completed the proof of Eq.~\eqref{reformulated_propoA2} in the special case with the MF model and BPR loss. 
    Then we will generalize our conclusion to other CF models with different loss functions. To achieve it, we just need to prove that the assumptions still hold with other CF models. It is easy to observe that Assumption~\ref{assumption:2} and Assumption~\ref{assumption:3} remain valid. As to Assumption~\ref{assumption:1}, the derivation is also similar to NCF and LightGCN since the partial derivative of $\mathcal{L}_{\text{BPR}}$ with respect to the embeddings of users and items are similar. 
    In the following, we prove that Assumption~\ref{assumption:1} still holds for the CF models with BCE loss function. Formally, we have 
    \begin{equation}
        \begin{aligned} 
        &\min_{\theta} \mathcal{L}_{\text{BCE}}(\mathcal{M}_\theta, \bm{D}) \\
        &=  \frac{1}{|\bm{P}|} \sum_{(u,i)\in\bm{P}}  \left[ -\log(\sigma(\mathcal{M}_\theta(u,i))) - \frac{1}{|\mathcal{I}|} \sum_{j\in\mathcal{I}} \log(1-\sigma(\mathcal{M}_\theta(u,j)))  \right]. 
        \end{aligned}
    \end{equation}
    Similarly, we could obtain the partial derivative of the loss function $\mathcal{L}_{\text{BCE}}$ with respect to the embeddings of user $u^*$ and item $i$ as Eq.~\eqref{equation:partial_derivative_u_bpr} and Eq.~\eqref{equation:partial_derivative_i_bpr}:
    \begin{equation}
    \begin{aligned}
        \frac{\partial{ \mathcal{L}_{\text{BCE}}(\mathcal{M}_\theta, \bm{D})}}{\partial{\bm{e}_{u^*}^\theta}} 
        &\approx \frac{1}{|\bm{P}|} \sum_{i\in \mathcal{N}_{u^*}} \phi_\theta(u^*,i,j) \cdot \bm{e}_i^\theta - \\
        & \frac{1}{|\bm{P}|} \frac{|\mathcal{N}_{u^*}|}{|\mathcal{I}|}  \sum_{j\in \mathcal{I}} \phi_\theta(u^*,i,j) \cdot \bm{e}_j^\theta, \\ 
        \frac{\partial{ \mathcal{L}_{\text{BCE}}(\mathcal{M}_\theta, \bm{D})}}{\partial{\bm{e}_{i^*}^\theta}} 
        &\approx \frac{1}{|\bm{P}|} \sum_{u\in \mathcal{N}_{i^*}} \phi_\theta(u,i^*,j) \cdot \bm{e}_u^\theta - \\
        & \frac{1}{|\bm{P}|}   \frac{1}{|\mathcal{I}|} \sum_{u' \in \mathcal{U}} |\mathcal{N}_{u'}| \cdot \phi_\theta(u',i,i^*) \cdot\bm{e}_{u'}^\theta,
    \end{aligned}
    \end{equation}
    where $\phi_\theta(u,i,j) = -(1-\sigma(\mathcal{M}_\theta(u,i)) = -\sigma(\mathcal{M}_\theta(u,j))$. Similarly, we can assume $\phi_\theta(u,i,j)$ approximately to become similar for different $(u, i, j)$ pairs and tends to be close to zero as training processes. Therefore, we can conclude that Assumption~\ref{assumption:1} also holds for the CF models (\eg MF, NCF, and LightGCN) with BCE loss. 
    Due to the popularity of BCE and BPR losses in recommender optimization, we mainly study these two loss functions and leave the exploration of more loss functions to future work. 
\end{proof}

\subsection{Theoretical Analysis of Proposition 2}\label{appendix:P_2_proofs}
\textbf{Reminder of Proposition 2.} The three-order path number between target user $u$ and target item $i$, $\bm{A}^3_{u,i}$, is equivalent to the weighted sum of the intermediate users who liked item $i$ in $\bm{A}$, where the weights are their interaction similarities with target user $u$, \ie the number of mutually liked items. 
\begin{proof}
If we denote all interactions of target item $i$ as $\bm{D}^*_r\in\{0,1\}^{|\mathcal{U}_r|\times 1}$, we can factorize $\bm{D}_r$ into $\left[\begin{array}{cc}
      \bar{\bm{D}}_r & \bm{D}^*_r \\
\end{array}\right]$, where $\bar{\bm{D}}_r\in\{0,1\}^{|\mathcal{U}_r|\times (|\mathcal{I}|-1)}$ represents the interactions on other items. In this way, the symmetric interaction matrix $\bm{A}=\left[\begin{array}{ccc}
      \bm{0} & \bar{\bm{D}}_r & \bm{D}^*_r \\
      (\bar{\bm{D}}_r)^T & \bm{0} & \bm{0} \\
      (\bm{D}^*_r)^T & \bm{0} & \bm{0} \\
\end{array}\right]$. Thereafter, the three-order path number from all users to the target item $i$ can be obtained from $\bm{A}^3 = \bm{A}\times\bm{A}\times\bm{A}$: $\bar{\bm{D}}_r(\bar{\bm{D}}_r)^T\bm{D}^*_r+\bm{D}^*_r(\bm{D}^*_r)^T\bm{D}^*_r$. 
Given a target user $u$, its three-order path number to the target item $i$ equals to a value in a row of $\bar{\bm{D}}_r(\bar{\bm{D}}_r)^T\bm{D}^*_r+\bm{D}^*_r(\bm{D}^*_r)^T\bm{D}^*_r$. 
Since there is no interaction between the target user $u$ and target item $i$ in $\bm{D}^*_r$, their corresponding value in $\bm{D}^*_r(\bm{D}^*_r)^T\bm{D}^*_r$ is also zero. As such, the three-order path number between $u$ and $i$ is only decided by $\bar{\bm{D}}_r(\bar{\bm{D}}_r)^T\bm{D}^*_r$, where $\bar{\bm{D}}_r(\bar{\bm{D}}_r)^T$ describes the interaction similarity between all users (\ie the number of mutually liked items) and $\bm{D}^*_r$ represents their preference over target item $i$. Afterward, the three-order path number from target user $u$ to target item $i$ is only affected by the users who liked item $i$ in $\bm{D}^*_r$ while the sum is weighted by the interaction similarity with target user $u$ via $\bar{\bm{D}}_r(\bar{\bm{D}}_r)^T$. Therefore, we have completed the proof of Proposition~\ref{pro:A3_path_user}.

\end{proof}

\vspace{-0.6cm}
\subsection{Dynamic Programming Algorithm}\label{appendix:DP}
\vspace{-0.1cm}
The budget-constrained treatment selection can be solved by using a dynamic programming approach for the knapsack problem, which is based on the principle of the optimal substructure. The dynamic programming algorithm efficiently computes the maximum value by iteratively solving smaller subproblems and integrating their solutions. 
Algorithm~\ref{algo:dp} presents pseudo-code for the process of calculating the optimal treatment after estimating the treatment effect of $Y^{\theta^*}_{u,i}(\bm{D}_f(t_u))$. In this algorithm, the maximal budget for each user $H$ means that we can vary different budgets $t_u \in \{0,1,2,...,H\}$ with $H\ll N$ to calculate $Y^{\theta^*}_{u,i}(\bm{D}_f(t_u))$.

\begin{algorithm}[t]
	\caption{Dynamic Programming Algorithm}  
	\label{algo:dp}
	\begin{algorithmic}[1]
		\Require a set of target users $\mathcal{U}_t$, treatment effects matrix $Y = \left[Y^{\theta^*}_{u,i}(\bm{D}_f(t_u)) \right]_{|\mathcal{U}_t| \times (H+1)}$ across different target users and budgets, the maximal budget number for each user $H$, and overall budget number $N$ for all target users. 
        \State \textbf{Function} dynamic\_programming($|\mathcal{U}_t|$, $Y$, $H$, $N$)
        \State $B$ = 1D array of size $H+1$ initialized with $\left[0,1,2,3,...,H\right]$;
        \State \algorithmiccomment{B contains the potential budget numbers of each user.}
        \State $dp,selected$ = 2D array of size $(|\mathcal{U}_t|+1) \times (N + 1)$ initialized with zeros;
        \State \algorithmiccomment{$selected$ is used to find the optimal allocation for each user.}
		\For {$i = 1$ to $|\mathcal{U}_t|$} 
                \For {$j = 0$ to $N$} 
                        \For {$k = 0$ to $H$} 
                                \If {$j \ge B[k]$} then
    		                          \State $dp[i][j]$ = $\max(dp[i][j],dp[i-1][j-B[k]]+Y[i][k])$;
    		                          \State $selected[i][j]$ = 1; 
		                      \EndIf
		              \EndFor
		      \EndFor
		\EndFor

        \State $P_{max}$ = $dp[|\mathcal{U}_t|][N]$;

        \State $T^*$ = 1D array of size $|\mathcal{U}_t|+1$ initialized with zeros;

        \State $j$ = $N$;
  
		\For {$i = |\mathcal{U}_t|$ down to $1$}
                
                \State $optimal\_budget$ = 0;
		    \For {$budget = 0$ to $H$} 
                        \If{$j \ge B[budget]$ and $selected[i][j]$ = 1} 
                                    \State $optimal\_budget$ = $budget$;
                                    \State \textbf{Break}
                        \EndIf
                        \State $T^*[i] \gets optimal\_budget$;
                        \State $j \gets j-B[optimal\_budget]$
                \EndFor
		\EndFor
        \State \textbf{Return} $P_{max}$, $T^*$; 
		\Ensure overall recommendation probability $P_{max}$, and the optimal budget allocation $T^*$.
	\end{algorithmic}
\end{algorithm}

\begin{table*}[h]
\renewcommand\arraystretch{0.8} 
\setlength{\abovecaptionskip}{0cm}
\setlength{\belowcaptionskip}{0cm}
\caption{Evaluation on three datasets of all users when overall fake user budget is 100. $*$ implies the improvements over the best baseline ``Target'' are statistically significant ($p$-value<0.05) under $t$-test. Due to the hit ratios on ML-1M and Yelp being quite small, we multiply all the hit ratios of ML-1M and Yelp by 10 and 100, respectively, for better comparison.}
\label{tab:main_all_user}
\begin{center}
\setlength{\tabcolsep}{2mm}{
\resizebox{\textwidth}{!}{
\begin{tabular}{lrrrrccccrrrr}
\toprule
 & \multicolumn{4}{c}{\textbf{ML-1M}} & \multicolumn{4}{c}{\textbf{Amazon}} & \multicolumn{4}{c}{\textbf{Yelp}} \\
 & \multicolumn{2}{c}{\textbf{Popular item}} & \multicolumn{2}{c}{\textbf{Unpopular item}} & \multicolumn{2}{c}{\textbf{Popular item}} & \multicolumn{2}{c}{\textbf{Unpopular item}} & \multicolumn{2}{c}{\textbf{Popular item}} & \multicolumn{2}{c}{\textbf{Unpopular item}} \\
 & \multicolumn{1}{c}{\textbf{HR@10}} & \multicolumn{1}{c}{\textbf{HR@20}} & \multicolumn{1}{c}{\textbf{HR@10}} & \multicolumn{1}{c}{\textbf{HR@20}} & \textbf{HR@10} & \textbf{HR@20} & \textbf{HR@10} & \textbf{HR@20} & \multicolumn{1}{c}{\textbf{HR@10}} & \multicolumn{1}{c}{\textbf{HR@20}} & \multicolumn{1}{c}{\textbf{HR@10}} & \multicolumn{1}{c}{\textbf{HR@20}} \\ \midrule
\multicolumn{1}{l}{\textbf{Before Attack}} & 0.050 & 0.109 & 0.000 & 0.000 & 0.002 & 0.003 & 0.000 & 0.000 & 0.000 & 0.000 & 0.000 & 0.000 \\
\multicolumn{1}{l}{\textbf{Random Attack}} & 0.050 & 0.082 & 0.000 & 0.000 & 0.018 & 0.197 & 0.013 & 0.043 & 0.008 & 0.016 & 0.024 & 0.024 \\
\multicolumn{1}{l}{\textbf{Segment Attack}} & 0.069 & 0.123 & 0.008 & 0.012 & 0.014 & 0.145 & 0.034 & 0.044 & 0.008 & 0.016 & 0.032 & 0.064 \\
\multicolumn{1}{l}{\textbf{Bandwagon Attack}} & 0.059 & 0.119 & 0.000 & 0.002 & 0.015 & 0.176 & 0.025 & 0.063 & 0.032 & 0.064 & 0.008 & 0.016 \\
\multicolumn{1}{l}{\textbf{Average Attack}} & 0.016 & 0.044 & 0.002 & 0.002 & 0.016 & 0.185 & 0.013 & 0.048 & 0.016 & 0.048 & 0.040 & 0.056 \\
\multicolumn{1}{l}{\textbf{WGAN}} & 0.041 & 0.076 & 0.012 & 0.012 & 0.019 & 0.254 & 0.034 & 0.076 & 0.024 & 0.032 & 0.040 & 0.040 \\
\multicolumn{1}{l}{\textbf{DADA-DICT}} & 0.096 & 0.156 & 0.034 & 0.062 & 0.039 & 0.079 & 0.103 & 0.204 & 0.096 & 0.144 & 0.072 & 0.096 \\ 
\multicolumn{1}{l}{\textbf{DADA-DIV}} & 0.094 & 0.132 & 0.039 & 0.057 & 0.044 & 0.082 & 0.85 & 0.211 & 0.089 & 0.138 & 0.077 & 0.099 \\ 
\multicolumn{1}{l}{\textbf{DADA}} & 0.122 & 0.194 & 0.050 & 0.084 & 0.092 & 0.144 & 0.103 & 0.204 & 0.084 & 0.145 & 0.076 & 0.158 \\ \midrule
\multicolumn{1}{l}{\textbf{AIA}} & 0.078 & 0.180 & 0.002 & 0.002 & 0.059 & 0.064 & 0.189 & 0.245 & 0.076 & 0.122 & 0.056 & 0.096 \\
\multicolumn{1}{l}{\textbf{\quad + Target}}& {\ul 0.146} & 0.262 & 0.072 & 0.111 & 0.062 & 0.077 & 0.249 & {\ul 0.301} & 0.075 & 0.194 & 0.048 & {\ul 0.144} \\
\multicolumn{1}{l}{\textbf{\quad + UBA(w/o $\mathcal{S}_\phi$)}} & \textbf{0.208} & {\ul 0.263} & {\ul 0.072} & \textbf{0.134} & {\ul 0.097} & {\ul 0.084} & {\ul 0.267 } & 0.292 & {\ul 0.099} & \textbf{0.245} & {\ul 0.048} & \textbf{0.160} \\
\multicolumn{1}{l}{\textbf{\quad + UBA(w/ $\mathcal{S}_\phi$)}} & {\ul 0.146} & \textbf{0.324} & \textbf{0.074} & {\ul 0.111} & \textbf{0.145} & \textbf{0.098} & \textbf{0.314} & \textbf{0.333} & \textbf{0.124} & {\ul 0.232} & \textbf{0.056} & 0.096 \\ \midrule
\multicolumn{1}{l}{\textbf{AUSH}}& 0.082 & 0.183 & 0.000 & 0.002 & 0.103 & 0.138 & 0.064 & 0.098 & 0.056 & 0.111 & 0.024 & 0.049 \\
\multicolumn{1}{l}{\textbf{\quad + Target}} & 0.146 & 0.260 & 0.062 & 0.082 & 0.176 & 0.378 & {\ul 0.082} & 0.095 & 0.048 & 0.120 & 0.056 & {\ul 0.168} \\
\multicolumn{1}{l}{\textbf{\quad + UBA(w/o $\mathcal{S}_\phi$)}} & {\ul 0.154} & {\ul 0.296} & \textbf{0.076} & \textbf{0.106} & {\ul 0.189} & {\ul 0.395} & \textbf{0.082} & {\ul 0.121} & \textbf{0.096} & {\ul 0.128} & {\ul 0.064} & 0.152 \\
\multicolumn{1}{l}{\textbf{\quad + UBA(w/ $\mathcal{S}_\phi$)}} & \textbf{0.190} & \textbf{0.371} & {\ul 0.067} & {\ul 0.092} & \textbf{0.189} & \textbf{0.423} & 0.071 & \textbf{0.189} &{ \ul 0.072 } & \textbf{0.128} & \textbf{0.064} & \textbf{0.184} \\ \midrule
\multicolumn{1}{l}{\textbf{Leg-UP}}  & 0.053 & 0.091 & 0.015 & 0.027 & 0.020 & 0.027 & 0.151 & 0.169 & 0.080 & 0.136 & 0.048 & 0.104 \\
\multicolumn{1}{l}{\textbf{\quad + Target}}  & 0.146 & {\ul 0.265} & 0.072 & {\ul 0.111} & {\ul 0.046} & 0.062 & 0.205 & 0.236 & 0.096 & 0.144 & 0.072 & 0.112 \\
\textbf{\quad + UBA(w/o $\mathcal{S}_\phi$)} & \textbf{0.196} & \textbf{0.342} & {\ul 0.081} & 0.101 & \textbf{0.049} & {\ul 0.064} & {\ul 0.221} & \textbf{0.266} & \textbf{0.112} & \textbf{0.160} & {\ul 0.112} & {\ul 0.168} \\
\textbf{\quad + UBA(w/ $\mathcal{S}_\phi$)} & {\ul 0.158} & 0.249 & \textbf{0.086} & \textbf{0.124} & 0.035 & \textbf{0.065} & \textbf{0.240} & {\ul 0.255} & {\ul 0.104} & {\ul 0.144 } & \textbf{0.128} & \textbf{0.192} \\ \bottomrule
\end{tabular}
}}
\end{center}
\vspace{-0.2cm}
\end{table*}

\begin{table}[h]
\setlength{\abovecaptionskip}{0cm}
\setlength{\belowcaptionskip}{0cm}
\caption{Performance of different backend models with UBA(w/o $S_{\phi}$) \wrt hyper-parameter tuning.}
\label{tab:para_alpha_beta}
\begin{center}
\setlength{\tabcolsep}{0.5mm}{
\resizebox{0.48\textwidth}{!}{

\begin{tabular}{ccccccc}
\hline
 & \multicolumn{2}{c}{\textbf{Leg-UP}} & \multicolumn{2}{c}{\textbf{AIA}} & \multicolumn{2}{c}{\textbf{AUSH}} \\
 & \textbf{HR@10} & \textbf{HR@20} & \textbf{HR@10} & \textbf{HR@20} & \textbf{HR@10} & \textbf{HR@20} \\ \hline
\textbf{$\alpha$= 1.0 $\beta$=1.0} & \textbf{0.18} & \textbf{0.32} & \textbf{0.24} & \textbf{0.4} & 0.28 & 0.40 \\
\textbf{$\alpha$= 0.5 $\beta$=1.0} & 0.16 & 0.24 & 0.22 & 0.36 & \textbf{0.30} & 0.36 \\
\textbf{$\alpha$= 1.0 $\beta$=0.3} & 0.18 & 0.30 & 0.24 & 0.38 & 0.26 & \textbf{0.42} \\ \hline
\end{tabular}

}
}
\end{center}
\vspace{-0.2cm}
\end{table}

\section{Experimental Settings}

\subsection{Hyper-parameter Tuning}\label{appendix:hyper-parameter}
In our UBA framework, we mainly introduce these hyper-parameters: fake user number $N$, the maximal fake user budget of each target user $H$, $\alpha$ and $\beta$ in $\alpha\cdot(\bm{A}^3_{u,i})^\beta$, and the proportion of accessible user interactions. We mainly follow precious studies~\cite{li2022revisiting,Lin_2022,lin2020attacking} for the selection of $H$ and $N$. As to $\alpha$ and $\beta$, they should be larger than zero to ensure positive correlations in Proposition~\ref{pro:A3_path_user}. We choose $\alpha$ and $\beta$ from $\{0.5,1\}$ and $\{0.3,1\}$, respectively. 
As shown in Table~\ref{tab:para_alpha_beta}, 
we find the UBA(w/o $\mathcal{S}_{\phi}$) is not very sensitive to $\alpha$ and $\beta$, and thus we choose {$\alpha=1.0$, $\beta = 1.0$} as our default parameters. Future work might consider more elaborate adjustments of $\alpha$ and $\beta$ in a larger scope. 
Lastly, we set the default proportion of user interactions accessible to attackers as 20\% and provide more results with varying proportions in Table~\ref{tab:data_ratio}. As to the hyper-parameters of the backend attackers and recommender models, we follow their original settings.

\subsection{Resource Costs}\label{appendix:resource}

All experiments are conducted on CentOS 1 machine with a 6-core Intel(R) Xeon(R) Gold 5218 CPU @ 2.30GHz, 1 NVIDIA GeForce RTX 3090 GPUs (24G), and 40G of RAM. 
In Table~\ref{tab:time}, we present the estimation time for the treatment effects, $Y^{\theta^*}_{u,i}(\bm{D}_f(t_u))$, with and without using $\mathcal{S}\phi$ on ML-1M. 
{The time costs of UBA involve four parts: 1) the estimation of treatment effects, 2) the budget allocation, 3) the fake user generation based on attacker models, and 4) the training of victim models. The whole attack process records the time of the fake user generation based on attacker models, and the training of victim models. One time estimation with $\mathcal{S}_\phi$ represents the time costs of generating fake users via attacker models and training the surrogate model.} For each process, we conduct ten experiments and calculate the average time to ensure reliability.



For each estimation using the surrogate model $\mathcal{S}\phi$, we can simultaneously obtain the recommendation probability of all target users under the same $t_u$, and the table shows the time used for such an estimation. 
Given limited overall fake user budgets $N$, the budgets a target user can obtain are limited ($H\ll N$). 
In our experiments, we usually use $H=6$ for 100 limited fake user budgets on 50 target users, where $H=6$ is sufficient to ensure a good attack performance. 
Since the estimation processes with different fake user budgets are independent, we can leverage multiple GPUs for parallel acceleration if necessary. 

We can also observe that the estimation time costs without $\mathcal{S}\phi$ are significantly lower. This is due to the high efficiency of calculating the three-order path numbers. 
By comparing the two estimation methods, we can find that: on one hand, using a reliable surrogate model results in more accurate effect estimation and better attack performance. However, training and attacking a surrogate model require more computational time during the estimation process. 
On the other hand, the estimation w/o $\mathcal{S}\phi$ calculates the three-order path numbers in $\bm{A}^3$, sacrificing little estimation accuracy while it is much faster in terms of computational time. 
In practical applications, these two estimation methods can be flexibly selected by the actual requirements.


   
\begin{table}[t]
\renewcommand\arraystretch{0.9} 
\setlength{\abovecaptionskip}{0cm}
\setlength{\belowcaptionskip}{0cm}
\caption{The table of time costs (minutes) on ML-1M.}
\label{tab:time}
\begin{center}
\setlength{\tabcolsep}{2.5mm}{

\begin{tabular}{llll}
\toprule
 \textbf{Methods} & \textbf{AIA} & \textbf{AUSH} & \textbf{Leg-UP} \\ \midrule
\textbf{One time estimation w/ $\mathcal{S}_\phi$} & 64.4m  & 52.1m  & 62.9m \\ 
\textbf{Estimation w/o $\mathcal{S}_\phi$} & 1.1m & 1.1m & 1.1m \\ 
\textbf{Whole attack process} & 50.6m  & 48.31m  & 69.3m \\ \bottomrule
\end{tabular}
}

\end{center}
\vspace{-0.2cm}
\end{table}

\begin{figure*}[t]
\setlength{\abovecaptionskip}{0cm}
\setlength{\belowcaptionskip}{-0.20cm}
  \centering
\includegraphics[scale=0.45]{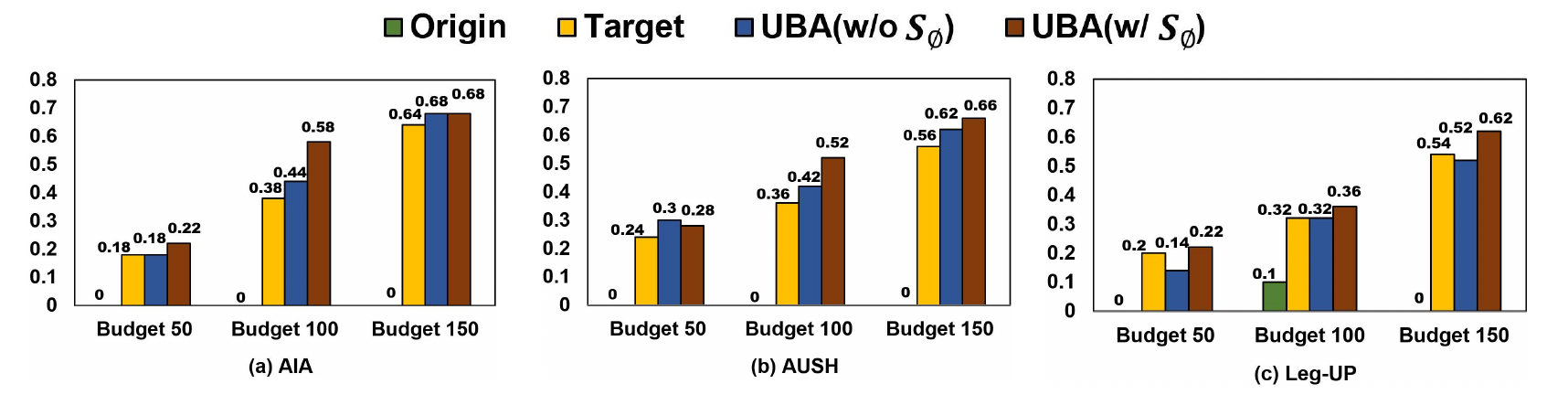}
  \hspace{-0.7in} 
  \caption{Performance comparison \wrt HR@10 under different attack budgets.}
  \label{fig:budget_unpop}
  \vspace{0cm}
\end{figure*}

\section{More Results}

\subsection{Overall Attack Performance on All Users}\label{appendix:overall_all_users}
In Table~\ref{tab:main_all_user}, we report the hit ratios on all users. Due to the different dataset sparsity, the hit ratios on the three datasets are not at the same magnitude. To better show the results, we multiply the hit ratios of ML-1M and Yelp by 10 and 100, respectively. 
Although the primary attack objective of target user attacks focuses on a specific set of users, UBA can also enhance the hit ratios of three backend attackers on all users, indicating that UBA not only achieves the SOTA attack performance on target users, but also guarantees the attack results on all users in the platform. In this light, UBA can also be utilized for injective attacks on all users, improving the attack performance of existing injective attackers.

\subsection{Attack Performance on All Users across Victim Models}\label{appendix:victim_all_users}

\begin{figure}[t]
\setlength{\abovecaptionskip}{0cm}
\setlength{\belowcaptionskip}{0cm}
  \centering 
  \subfigure{
    \includegraphics[scale=0.42]{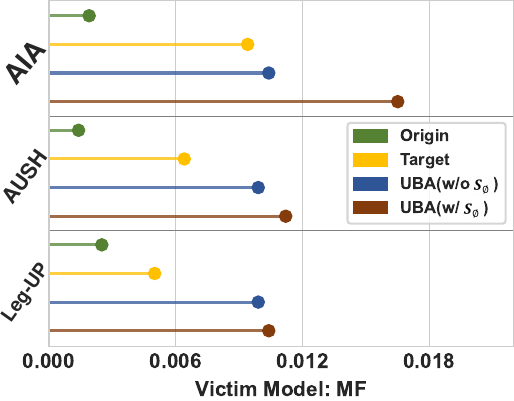}} 
  \hspace{0.2in}
  \subfigure{
    \includegraphics[scale=0.42]{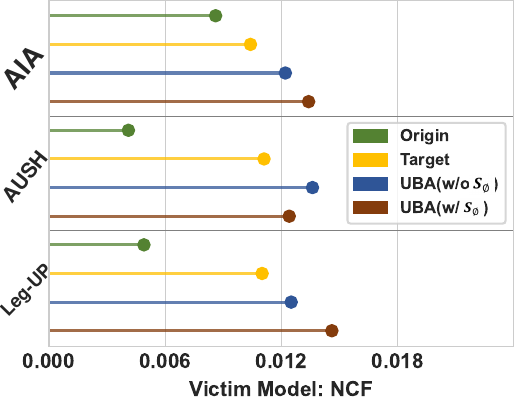}} 
  \caption{Generalization of UBA \wrt HR@10 across different victim models on all users.}
  \label{fig:generation_all_users}
  \vspace{0cm}
\end{figure}

Figure~\ref{fig:generation_all_users} visualizes the attack results of using MF and NCF as victim models on ML-1M. By inspecting Figure~\ref{fig:generation_all_users} and the results of LightGCN in Table~\ref{tab:main_all_user}, we can have the observations that UBA shows higher hit ratios than ``Origin'' and ``Target'' on all users, indicating good generalization ability of UBA across different victim models even for all users.

\subsection{Attack Performance on All Users with Defense Models}\label{appendix:detector_all_users}

When defense models are deployed, we present the attack results of three backend attackers with ``Target'' and UBA on all users of ML-1M in Table~\ref{tab:detector_allusers}. 
From the table, we find 1) the detectors are still effective in most cases although they are not quite robust, and 2) UBA methods can still have superior attack performance than the baselines when the detectors are applied. 


\begin{table}[h]
\setlength{\abovecaptionskip}{0cm}
\setlength{\belowcaptionskip}{0cm}
\caption{Attack performance comparison w/ and w/o using defense models on all users.}
\label{tab:detector_allusers}
\begin{center}
\setlength{\tabcolsep}{1.5mm}{
\resizebox{0.48\textwidth}{!}{
\begin{tabular}{cccccccccc}
\toprule
\multicolumn{1}{l}{} & \multicolumn{1}{l}{} & \multicolumn{2}{c}{\textbf{AIA}} & \multicolumn{2}{c}{\textbf{+ Target}} & \multicolumn{2}{c}{\textbf{+ UBA(w/o $\mathcal{S}_\phi$)}} & \multicolumn{2}{c}{\textbf{+ UBA(w/ $\mathcal{S}_\phi$)}} \\
 &  & \textbf{Origin} & \textbf{Detector} & \textbf{Origin} & \textbf{Detector} & \textbf{Origin} & \textbf{Detector} & \textbf{Origin} & \textbf{Detector} \\ \midrule
 & \multicolumn{1}{c}{\textbf{HR@10}} & 0.0078 & 0.0087 & 0.0146 & 0.0132 & 0.0208 & 0.0182 & 0.0146 & 0.0141 \\
\multirow{-2}{*}{\textbf{PCA}} & \multicolumn{1}{c}{\textbf{HR@20}} & 0.0180 & 0.0192 & 0.0262 & 0.0251 & 0.0324 & 0.0235 & 0.0263 & 0.0240 \\ 
 & \multicolumn{1}{c}{\textbf{HR@10}} & 0.0078 & 0.0087 & 0.0146 & 0.0132 & 0.0208 & 0.0182 & 0.0146 & 0.0141 \\
\multirow{-2}{*}{\textbf{FAP}} & \multicolumn{1}{c}{\textbf{HR@20}} & 0.0180 & 0.0192 & 0.0262 & 0.0251 & 0.0324 & 0.0235 & 0.0263 & 0.0240 \\ \midrule
\multicolumn{1}{l}{} & \multicolumn{1}{l}{} & \multicolumn{2}{c}{\textbf{AUSH}} & \multicolumn{2}{c}{\textbf{+ Target}} & \multicolumn{2}{c}{\textbf{+ UBA(w/o $\mathcal{S}_\phi$)}} & \multicolumn{2}{c}{\textbf{+ UBA(w/ $\mathcal{S}_\phi$)}} \\ \midrule
 & \multicolumn{1}{c}{\textbf{HR@10}} & 0.0082 & 0.0107 & 0.0146 & 0.0142 & 0.0154 & 0.0149 & 0.0168 & 0.0210 \\
\multirow{-2}{*}{\textbf{PCA}} & \multicolumn{1}{c}{\textbf{HR@20}} & 0.0183 & 0.0203 & 0.0260 & 0.0260 & 0.0296 & 0.0281 & 0.0287 & 0.0336 \\ 
 & \multicolumn{1}{c}{\textbf{HR@10}} & 0.0082 & 0.0068 & 0.0146 & 0.0130 & 0.0154 & 0.0084 & 0.0168 & 0.0093 \\
\multirow{-2}{*}{\textbf{FAP}} & \multicolumn{1}{c}{\textbf{HR@20}} & 0.0183 & 0.0125 & 0.0260 & 0.0228 & 0.0296 & 0.0172 & 0.0287 & 0.0169 \\ \midrule
 & \multicolumn{1}{c}{} & \multicolumn{2}{c}{\textbf{Leg-UP}} & \multicolumn{2}{c}{\textbf{+ Target}} & \multicolumn{2}{c}{\textbf{+ UBA(w/o $\mathcal{S}_\phi$)}} & \multicolumn{2}{c}{\textbf{+ UBA(w/ $\mathcal{S}_\phi$)}} \\ \midrule
 & \multicolumn{1}{c}{\textbf{HR@10}} & 0.0053 & 0.0025 & 0.0146 & 0.0091 & 0.0196 & 0.0146 & 0.0158 & 0.0160 \\
\multirow{-2}{*}{\textbf{PCA}} & \multicolumn{1}{c}{\textbf{HR@20}} & 0.0091 & 0.0044 & 0.0265 & 0.0181 & 0.0342 & 0.0258 & 0.0249 & 0.0263 \\ 
 & \multicolumn{1}{c}{\textbf{HR@10}} & 0.0053 & 0.0025 & 0.0146 & 0.0089 & 0.0196 & 0.0091 & 0.0158 & 0.0130 \\
\multirow{-2}{*}{\textbf{FAP}} & \multicolumn{1}{c}{\textbf{HR@20}} & 0.0091 & 0.0044 & 0.0265 & 0.0160 & 0.0342 & 0.0254 & 0.0249 & 0.0228 \\ \bottomrule
\end{tabular}
}
}
\end{center}
\vspace{-0.2cm}
\end{table}

\subsection{Attack of Unpopular Items with Varying Budgets}\label{appendix:unpopular_budgets}

In Figure~\ref{fig:budget_unpop}, we present the attack performance with the unpopular target items in ML-1M \wrt varying fake user budgets. We can find that the trends on the unpopular items are similar to those on the popular items presented in Figure~\ref{fig:budget}. 

For popular or unpopular items, both UBA w/ and w/o $\mathcal{S}_{\phi}$ achieve better attack performance than the backend attackers and ``Target'' under different budgets. This further verifies the robustness of UBA \wrt different fake user budgets and the target items with different popularity.

\section{Related Work}\label{appendix:related_work}
\subsection{Uplift Modeling}
Uplift, a term commonly used in marketing, refers to the disparity in purchasing behaviors between customers who receive a promotional offer (the treated group) and those who do not (the control group)~\cite{zhang_unified_2021,liu2022towards}. In causal terms, uplift essentially measures the causal effect of a treatment~\cite{Ai-LBCF-2022, sun2022counterfactual}, such as a promotion, on the desired outcome, such as customer purchasing behaviors. While uplift modeling has been extensively researched in the fields of machine learning and marketing\cite{Causal-gutierrez-2017, Ai-LBCF-2022, Tu-Personalized-2021}, its application in the realm of recommendation systems has received relatively little attention~\cite{xie21causCF,yao21survey,wu2022opportunity}.

Initial studies in this area have primarily focused on exploring the potential of uplift modeling to regulate the exposure proportion of different item categories~\cite{yu-MDP2-2022}. However, in this work, we take a different perspective by defining the assigning of fake user budgets in injective attacks as the treatment variable. We estimate the difference in recommendation probabilities of the target item on target users as the uplifts. Leveraging these estimated uplifts, our objective is to identify the optimal treatment strategy that maximizes the overall recommendation probabilities for all target users.

Target user attacks usually have limited fake user budgets, which can be formulated as a budget-constrained optimization problem. Therefore, the core is how to maximize the attack performance by using limited budgets. 
By adopting uplift modeling techniques in the context of recommender attacks, we can optimally utilize the limited budgets to increase their causal effects.

\subsection{Injective Attacks}
Injective attacks aim to disrupt the recommendation strategy of a victim recommender model in order to increase the exposure of a target item to all users~\cite{o2005recommender,mobasher2007toward,deldjoo2019assessing,you23anti}. To achieve this, attackers often inject fake users into the training data of the victim model for interference. As such, the core of injective attacks is to construct the interaction behaviors of fake users.

The construction methods of fake user interactions can be mainly categorized into three types: 
\begin{itemize}[leftmargin=*]
    \item Heuristic attackers create fake users based on some heuristic rules. The goal is to enhance the similarity between fake users and real users while improving the co-occurrence of the target item and some selected items, thus increasing the exposure probability of the target item. Existing methods, such as Random Attack, Average Attack, Bandwagon Attack, and Segment Attack, usually leverage different human-designed rules to determine the selected items~\cite{1167344,burke2005limited,kaur2016shilling,1565730}. 
    
    \item Gradient attacks optimize the attack objective in a continuous space to directly adjust the fake user interactions, and then truncate continuous values into discrete fake user interactions~\cite{li2016data,fang2020influence,yang2017fake,zhang2020practical,fang2018poisoning,huang2021data}.   
    \item Neural attacks leverage generative neural networks to generate fake user interactions. To maximize the attack objective, various methods, such as WGAN, AIA, AUSH, and Leg-UP, have been proposed to improve the effectiveness of neural attackers~\cite{https://doi.org/10.48550/arxiv.1701.07875,tang2020revisiting,lin2020attacking,Lin_2022}. Besides, reinforcement learning is also employed when the attacker can rely on some sparse feedback from the victim model~\cite{zhang2022targeted,fan2021attacking,song2020poisonrec}.
\end{itemize}

Data security and privacy concerns are making the assumption of attack knowledge increasingly important. Attack knowledge assumptions can be categorized into white-box, grey-box, and black-box settings. White-box settings assume that the structure and parameters of the victim recommender model, as well as all user-item interaction data, are accessible~\cite{fang2018poisoning,fang2020influence,li2016data}. Besides, grey-box settings have limited access to the user-item interactions and knowledge about the victim model~\cite{yang2017fake,cao2020adversarial,wu2021fight}. Lastly, black-box settings can only have a few feedback from the spy users~\cite{burke2005limited,chen22ke, yue21bbattack}. In this work, we utilize a reasonable setting, where only the interactions of partial users are accessible to the attackers and the victim models are totally unknown. 

\subsection{Defense Methods}\label{appendix:related_defense}

A defense model plays a crucial role in preventing fake users from disturbing victim models. Existing methods can be roughly classified into two categories. One direction is to design more robust recommender models against injective attacks~\cite{wu21Fight,wu21triple}. However, this line of methods cannot be applied to current well-designed industry recommender models. To complement this, some studies focus on detecting and excluding fake users from the training data of recommender models~\cite{aktukmak2019quick,bhaumik2006securing}. 

Based on whether the detection methods use the labels of fake users for training, current direction methods can be categorized into three types: supervised models~\cite{dou2018collaborative,li2016shilling,yang2018detection}, semi-supervised models~\cite{cao2013shilling, wu2011semi}, and unsupervised models~\cite{mehta2009unsupervised,zhou2016svm,davoudi2017detection,10.1145/3442381.3449813, you23anti}. Technically speaking, these methods have explored sequential GANs~\cite{DBLP:journals/corr/abs-2012-02509}, RNNs~\cite{gao2020shilling}, and CNNs~\cite{9498095} for detection. In future work, these methods can be evaluated by comparing the performance in defending against target user attackers. 

In addition, a more critical direction is adversarial training~\cite{tang2019adversarial, he2018adversarial, liu2020certifiable}. The defense models can be trained with some transparent attackers via multi-turn attack and defense, enhancing the robustness and generalization of defense models in the presence of fake user data. Our UBA framework can be applied to improve extensive attackers, serving as a great assistant for such adversarial training.


\end{document}